\documentclass[%
 reprint,
superscriptaddress,
 amsmath,amssymb,
 aps,
]{revtex4-2}

\usepackage{graphicx}
\usepackage{dcolumn}
\usepackage{bm}

\usepackage{pdfsync}
\usepackage{amsmath}    
\usepackage{amsfonts} 
\usepackage{amssymb}
\usepackage{MnSymbol}
\usepackage{mathrsfs} 
\usepackage{comment}

\usepackage{textcomp} 
\usepackage{bbm}
\usepackage{float}    
\usepackage{accents} 
\usepackage{natbib}
\usepackage{graphicx}   
\usepackage[caption=false]{subfig}
\usepackage{mathtools}
\usepackage{graphics}
\usepackage{epsfig}
\usepackage{multirow}
\usepackage{bibentry}
\usepackage{braket}
\usepackage{algorithm}
\usepackage{algorithmic}
\usepackage{listings}
\usepackage{setspace}
\usepackage{fancyhdr}
\usepackage{slashed}
\usepackage{xfrac}
\usepackage{cases}
\usepackage[retainorgcmds]{IEEEtrantools}
\usepackage[pdftex, citecolor=blue, urlcolor=blue, linkcolor=blue, colorlinks=true, bookmarksopen=true]{hyperref}
\hypersetup{colorlinks,allcolors=black}
\allowdisplaybreaks  

\usepackage[table]{xcolor}

\newcommand{\pT}{\hat{+}}
\newcommand{\mT}{\hat{-}}
\newcommand{\muT}{\hat{\mu}}

\newcommand{\itP}[1]{\hat{#1}}

\begin{document}

\preprint{APS/123-QED}

\title{Interpolating conformal algebra in \texorpdfstring{$(1+1)$}{TEXT} dimensions between the instant form and the light-front form of relativistic dynamics}

\author{Chueng-Ryong Ji}
\affiliation{Department of Physics, North Carolina State University, Raleigh, North Carolina 27695-8202, USA}

\author{Hariprashad Ravikumar}
\affiliation{Department of Physics, New Mexico State University, Las Cruces, New Mexico 88003-8001, USA}

\begin{abstract}
We present the interpolating conformal algebra between the instant form dynamics (IFD) and the light-front dynamics (LFD) in $(1+1)$ dimensions, along with a $4\times4$ interpolating projective spacetime matrix representation. While there are six generators in the $(1+1)$ dimensional conformal algebra, the number of kinematic and dynamic generators dramatically changes in LFD, maximizing (minimizing) the number of kinematic (dynamic) generators to four (two) with respect to two (four) kinematic (dynamic) generators in IFD, as well as in any other forms of dynamics between IFD and LFD. It confirms and signifies the utility of LFD, saving substantial dynamical efforts in solving the $(1+1)$ dimensional quantum field theories. We also present $2\times2$ Pauli matrix representation of $(1+0)$ and $(0+1)$ conformal groups, and creation/annihilation operators of quantum simple harmonic oscillator representations of $(1+0)$ dimensional conformal groups.

\end{abstract}
\keywords{conformal algebra, light-front dynamics, instant form dynamics, Witt algebra}
\maketitle

\section{Introduction}
\label{sec:introduction}

It is essential to study the conformal symmetry in discussing the fundamental properties of spacetime in analyzing the relativistic quantum field theories~\cite{Wess1960, Kastrup1966, SalamMack1969, Gross1970} as well as developing characteristics of duality as exemplified in AdS/CFT correspondence~\cite {Weinberg_2010, Brodsky_2015, Maldacena2016}. In the early 20th century, Cunningham  \cite{Cunningham1910} and Bateman  \cite{Bateman1910} demonstrated that the Maxwell equations are invariant under conformal transformations. Later, Dirac  \cite{Dirac1936} proved that the massless version of his well-known equation in relativistic quantum mechanics also exhibits invariance under conformal transformations. In the 1960s, Gross \cite{Gross1970} explored the scale invariance as an asymptotic (high-energy) symmetry of scattering amplitudes, demonstrating that the scale invariance of Lagrangian field theories implies conformal invariance. A partial list of literature discussing the physical significance and potential applications of scale and conformal invariance can be found in Refs. \cite{Wess1960, Gürsey1956, Fulton1962, Kastrup1966, SalamMack1969, Wilson1969, Gross1970}. Additionally, a historical review in Ref. \cite{Kastrup2008} provides further early references on this topic. Aspects of the conformal group in two-dimensional Minkowski spacetime, with representations of the algebra and of the group within a field-theoretic Schrodinger representation for bosons and fermions, were reviewed in Ref. \cite{Jackiw1990}.

In this work, we start by discussing the conformal symmetry in the lowest dimensions, namely $(0+1)$ and $(1+0)$ dimensions, and then move on to discuss the conformal symmetry in 
$(1+1)$ dimensions. In particular, we elaborate the $(1+1)$ dimensional conformal algebra with the interpolation between the two different forms of relativistic dynamics, i.e., the instant form dynamics (IFD) and the light-front dynamics (LFD). With this interpolation, we discuss a dramatic difference in the number of kinematic generators that leave the time defined in the given form of the dynamics invariant
between IFD and LFD. Figuring out which generator is kinematic in what form of dynamics is useful as the more kinematic generators save the correspondingly more dynamic efforts in solving the relativistic quantum field theories, such as QED and QCD~ \cite{ji2023relativistic}.

In general, the conformal transformation $x\longmapsto x'$ in $d$ dimension can be defined as \cite{Francesco,Blumenhagen}, 
\begin{align}
    \frac{\partial x'^{\alpha}}{\partial x^{\mu}}\frac{\partial x'^{\beta}}{\partial x^{\nu}}g'_{\alpha\beta}=\Lambda(x)g_{\mu\nu},
\end{align}
meaning the metric is preserved up to a scale factor $\Lambda(x)$ under conformal transformation, where $\mu,\nu\in\{0,d-1\}$. The scale $\Lambda(x)=1$ corresponds to the Poincar\'e group consisting of translations, rotations, and Lorentz transformations. Consider an infinitesimal transformation, $x'^{\mu}=x^{\mu}+\epsilon^{\mu}(x)+\mathcal{O}(\epsilon^2)$, then the metric changes by, $\delta g_{\mu\nu}=\partial_{\mu}\epsilon_{\nu}(x)+\partial_{\nu}\epsilon_{\mu}(x)$. Conformality condition then requires, 
\begin{align}
    \partial_{\mu}\epsilon_{\nu}(x)+\partial_{\nu}\epsilon_{\mu}(x)=F(x)\delta_{\mu\nu},\label{Killing}
\end{align}
where $F(x)=\Lambda(x)-1-\mathcal{O}(\epsilon^2)$. The Eq.\eqref{Killing} is called the conformal Killing equation. Contraction with $\delta^{\mu\nu}$ yields $F(x)=\frac{2}{d}\partial_{\mu}\epsilon^{\mu}$. There are only 4 classes of solutions for $\epsilon_{\mu}(x)$ which are, $\epsilon^{\mu}(x)=a^{\mu}$ (infinitesimal translations), $\epsilon^{\mu}(x)=M^{\mu}_{~\nu}x^{\nu}$ (infinitesimal rotations and Lorentz transformations), $\epsilon^{\mu}(x)=\lambda x^{\mu}$ (infinitesimal scaling), and $\epsilon^{\mu}(x)=2(b.x) x^{\mu}-x^2b^{\mu}$ (Infinitesimal special conformal transformations or SCT). The generators of conformal transformations are: 
\begin{align}
    P_{{\mu}}=&i\partial_{{\mu}}&& \text{(translation)}\\
    M_{{\mu}{\nu}}=&i\left(x_{{\mu}}\partial_{{\nu}}-x_{{\nu}}\partial_{{\mu}}\right)&& \text{(rotation)}\\
    D=&ix_{{\mu}}\partial^{{\mu}}&& \text{(dilation)}\\
    \mathfrak{K}_{{\mu}}=&i\left(2x_{{\mu}}x_{{\nu}}\partial^{{\nu}}-x^2\partial_{{\mu}}\right)&& \text{(SCT)}
\end{align}

In finite form, the SCT will be $x'^{\mu }={\frac {x^{\mu }-b^{\mu }x^{2}}{1-2b\cdot x+b^{2}x^{2}}}$, this can be understood as an inversion of $x^\mu$, followed by a translation $b^\mu$, and followed again by an inversion \cite{Blumenhagen}.

In the next section, Sec.~\ref{sec:conformal0110}, we present the one-dimensional conformal algebra and its representation in terms of creation and annihilation operators of a harmonic oscillator. In Sec .~\ref{sec:conformal}, we extend the interpolation method from the Poincar\'e group to the conformal group. We demonstrate that the $SO(2, 1 + 1)$ algebra decomposes into a direct sum of two identical algebras in the light-front limit. The interpolating Witt-type algebra is introduced in Sec.~\ref{sec_Witt-type}. 
In Sec.~\ref{sec:projectivespace}, we show the projective space representation and discuss kinematic and dynamic generators in various interpolation angles, along with the $4\times4$ matrix representation of all conformal generators. 
Summary and conclusion follow in the last section, Sec.~\ref{summary-conclusion}. Finally, in Appendix~\ref{Dilated-vacuum},  we present the more detailed derivation discussed in Sec.~\ref{sec:conformal0110} regarding the coherent vacuum ($\ket{\Omega_{D_{3}}}$) associated with the dilation.

\section{Conformal Algebra in \texorpdfstring{$(0+1)$}
{Lg} and \texorpdfstring{$(1+0)$}{Lg} Dimension}
\label{sec:conformal0110}
 In 0-space \& 1-time $(0+1)$ dimension, as the only variable is time $x^0$, we may consider three operations with time, namely evolution (i.e. translation of time $x^{0\prime}=x^0+c$), scaling (i.e. $x^{0\prime} = \lambda x^0$) and the translation of inversed-time (i.e. $\frac{1}{x^{0\prime}}=\frac{1}{x^0}+\frac{1}{b}$). The generators of these three operations correspond to the three conformal generators, namely
\begin{align}
    P^{(0+1)}_{0}=&i\partial_{0},\label{P01}\\
    D^{(0+1)}_{0}=&ix_{0}\partial_{0},\label{D01}\\
    \mathfrak{K}^{(0+1)}_{{0}}=&ix_{0}x_{0}\partial_{{0}}.\label{KK01}
\end{align}
The transformations generated by $(0+1)$-conformal generators can be embodied in the single projective transformation (Möbius transformation) defined as
\begin{align}
    x_{0}\rightarrow x^{\prime}_{0}=\frac{\alpha x_{0} +\beta}{\gamma x_{0} +\delta},
\end{align}
where the real numbers: $\alpha, \beta, \gamma, \delta$ can be considered as the elements of a real unimodular matrix, namely $\begin{pmatrix}
    \alpha &  \beta \\
   \gamma & \delta
\end{pmatrix}$ with its determinant equal to 1, i.e., $\alpha\delta-\beta\gamma=1$. The corresponding Pauli matrix representation for $(0+1)$ conformal algebra was explored~ \cite{Fubini1976}, which may be summarized as 
\begin{align}
    P^{(0+1)}_{0}=&i\sigma_{-}=\begin{pmatrix}
        0&0\\
        i&0
    \end{pmatrix},\label{P01Pauli}\\
    D^{(0+1)}_{0}=&\frac{i}{2}\sigma_{3}=\begin{pmatrix}
        \frac{i}{2}&0\\
        0&-\frac{i}{2}
    \end{pmatrix},\label{D01Pauli}\\
    \mathfrak{K}^{(0+1)}_{{0}}=&-i\sigma_{+}=\begin{pmatrix}
        0&-i\\
        0&0
    \end{pmatrix},\label{KK01Pauli}
\end{align}
where $\sigma_{\pm}=\frac{\sigma_{1}\pm i\sigma_{2}}{2}$. With this representation, it is easy to check that the (0+1)-conformal group is isomorphic to the $SO(2,1)$ group. The commutation relations among the three generators of the conformal symmetry group can be found as shown in Table~\ref{tabeIFD01}.  
\begin{table}[h!]
\centering
\caption{\label{tabeIFD01}$0+1$ conformal algebra in IFD}
\scalebox{1.2}{
\begin{tabular}{|>{\centering\arraybackslash}p{1.5cm}||>{\centering\arraybackslash}p{1.5cm}|>{\centering\arraybackslash}p{1.5cm}|>{\centering\arraybackslash}p{1.5cm}|}
        \hline
        \rule{0pt}{16pt} & $\mathfrak{K}^{(0+1)}_{0}$ &$P^{(0+1)}_{0}$ &  $D^{(0+1)}_{0}$ \\
        \hline\hline
        $\mathfrak{K}^{(0+1)}_0$ &  0                  &$-2iD^{(0+1)}_{0}$ & $-i\mathfrak{K}^{(0+1)}_{0}$ \\
        \hline
        $P^{(0+1)}_0$            &  $2iD^{(0+1)}_{0}$            &0        & $iP^{(0+1)}_{0}$             \\
        \hline
        $D^{(0+1)}_0$            &  $i\mathfrak{K}^{(0+1)}_{0}$ &$-iP^{(0+1)}_{0}$ & 0                    \\
        \hline
      \end{tabular}}
\end{table}

Such a derivation of the $(0+1)$-conformal algebra may be regarded as the simplest possible conformal symmetry built in the lowest dimension. Nevertheless, it has been explored for the local isomorphism between this lowest dimension $(0+1)$-conformal group and the isometries of AdS$_2$ represented by the group $SO(2, 1) $for the discussion of confinement in the light-front Hamiltonian dynamics~ \cite{Brodsky_2015}. The 1-space \& 0-time $(1+0)$ dimensional conformal algebra may be derived correspondingly following the procedure illustrated in the above derivation of the 0-space \& 1-time $(0+1)$ dimensional conformal algebra. Namely, the correspondence between $x^0$ and $x^3$ provides the transformation generated by $(1+0)$ conformal generators, and thus the space $x^3$ transformation can also be embodied as a single projective transformation (Möbius transformation) defined as
\begin{align}
    x_{3}\rightarrow x^{\prime}_{3}=\frac{\alpha x_{3} +\beta}{\gamma x_{3} +\delta},
\end{align}
where the real numbers: $\alpha, \beta, \gamma, \delta$ can be considered as the elements of a real unimodular matrix, namely $\begin{pmatrix}
    \alpha &   \beta \\
  \gamma & \delta
\end{pmatrix}$ with $\alpha\delta-\beta\gamma=1$. 
The three conformal generators here may also be identified 
correspondingly to Eqs.\eqref{P01}-\eqref{KK01} as 
\begin{align}
    P^{(1+0)}_{3}=&i\partial_{3},\label{P10}\\
    D^{(1+0)}_{3}=&-ix_{3}\partial_{3}.\label{D10}\\
    \mathfrak{K}^{(1+0)}_{{3}}=&-ix_{3}x_{3}\partial_{{3}},\label{KK10}
\end{align}
respectively. Following the Pauli-matrix representation of the $(0+1)$ conformal algebra, we may correspond the three generators in Eqs.\eqref{P01Pauli}-\eqref{KK01Pauli} as 
\begin{align}
    P^{(1+0)}_{3}=&i\sigma_{-}=\begin{pmatrix}
        0&0\\
        i&0
    \end{pmatrix},\label{P10Pauli}\\
    D^{(1+0)}_{3}=&\frac{i}{2}\sigma_{3}=\begin{pmatrix}
        \frac{i}{2}&0\\
        0&-\frac{i}{2}
    \end{pmatrix},\label{D10Pauli}\\
    \mathfrak{K}^{(1+0)}_{{3}}=&+i\sigma_{+}=\begin{pmatrix}
        0&i\\
        0&0
    \end{pmatrix},\label{KK10Pauli}
\end{align}
where $\sigma_{\pm}=\frac{\sigma_{1}\pm i\sigma_{2}}{2}$. 
It yields again the isomorphic $SO(2,1)$ group representation of the $(1+0)$ conformal algebra as we expect from the correspondence between space $x^3$ and time $x^0$. The commutation relations among the three generators Eqs.\eqref{P10Pauli}-\eqref{KK10Pauli} can also be summarized correspondingly to Table~\ref{tabeIFD01} as 
Table~\ref{tabeIFD10}.   

\begin{table}[h!]
\centering
\caption{\label{tabeIFD10}$1+0$ conformal algebra in IFD}
\scalebox{1.2}{
\begin{tabular}{|>{\centering\arraybackslash}p{1.5cm}||>{\centering\arraybackslash}p{1.5cm}|>{\centering\arraybackslash}p{1.5cm}|>{\centering\arraybackslash}p{1.5cm}|}
        \hline
        \rule{0pt}{16pt} &  $\mathfrak{K}^{(1+0)}_{3}$ & $P^{(1+0)}_{3}$ &$D^{(1+0)}_{3}$ \\
        \hline\hline
        $\mathfrak{K}^{(1+0)}_3$ &  0                   &$2iD^{(1+0)}_{3}$  & $-i\mathfrak{K}^{(1+0)}_{3}$ \\
        \hline
        $P^{(1+0)}_3$            &  $-2iD^{(1+0)}_{3}$           &0         & $iP^{(1+0)}_{3}$            \\
        \hline
        $D^{(1+0)}_3$            &  $i\mathfrak{K}^{(1+0)}_{3}$ & $-iP^{(1+0)}_{3}$  &0                   \\
        \hline
      \end{tabular}}
\end{table}

While we achieve essentially identical conformal algebras for $(0+1)$ dimensions and $(1+0)$ dimensions, we note a remarkable difference between time $x^0$ and position $x^3$, as time is absolute and doesn't mix with position in non-relativistic quantum mechanics. Indeed, there is a theorem known as Pauli's theorem \cite{Pauli1980}, which states that time cannot be an operator but should be taken as a parameter, more precisely, an evolution parameter of a dynamical system.   

However, the position can be taken as an operator in quantum mechanics, and the $(1+0)$-conformal algebra summarized in Table~\ref{tabeIFD10} can be re-derived from the perspectives of the quantum operator algebra. Namely, the exact conformal algebra in $(1+0)$ dimension can be re-derived by taking the position operator in the matrix formulation of the one-dimensional simple harmonic oscillator (SHO) ($\hbar=\omega=c=1$ unit is taken here), with $x_3=\frac{a+a^{\dagger}}{\sqrt{2}}$ and $P_3=i\frac{a^{\dagger}-a}{\sqrt{2}}$.
Thus, we may utilize the creation-annihilation operators and discuss the (1+0)-conformal algebra in terms of the ladder operators in the Hilbert space, which can be defined by the number operator $N=a^{\dagger} a$, satisfying $[N, a] = -a$ and $[N,a^{\dagger}]= a^\dagger$ (where, $[A,B]=AB-BA$). 

Since the generators of dilation and SCT can be reduced to $D_{3}=\frac{-1}{2}\left(\{x_{3}, P_{3}\}+i\right)$ and $\mathfrak{K}_{{3}}=\frac{-1}{2}\left(\{x_{3}x_{3}, P_{3}\}+i2x_{3}\right)$, (where, $\{A,B\}=AB+BA$) we find
\begin{align}
    P^{(1+0)}_{3}=&i\frac{(a^{\dagger}-a)}{\sqrt{2}}\label{creation_anni_P},\\
    D^{(1+0)}_{3}=&\frac{-i}{2}\left(a^{\dagger 2} - a^2+1\right)\label{creation_anni_D},\\
    \mathfrak{K}^{(1+0)}_{{3}}=&\frac{-i}{2\sqrt{2}}\left(3a+3a^{\dagger}+  a^{\dagger 2}a - a^3 +  a^{\dagger 3} - a^2 a^{\dagger}\right)\label{creation_anni_KK}.
\end{align}

Both Pauli matrices $SO(2,1)$ representation, Eqs. \eqref{P10Pauli}-\eqref{KK10Pauli} and harmonic oscillator representation Eqs. \eqref{creation_anni_P}-\eqref{creation_anni_KK} obey the $(1+0)$ conformal algebra stated in Table \ref{tabeIFD10}. From the correspondence between them, we note mapping both representations as
\begin{align}
    i\sigma_{-}&\leftrightarrow i\frac{(a^{\dagger}-a)}{\sqrt{2}},\\
    i\sigma_{+}&\leftrightarrow \frac{-i}{2\sqrt{2}}\left(3a+3a^{\dagger}+  a^{\dagger 2}a - a^3 +  a^{\dagger 3} - a^2 a^{\dagger}\right),\\
    \frac{i}{2}\sigma_{3}&\leftrightarrow \frac{-i}{2}\left(a^{\dagger 2} - a^2+1\right).
\end{align} 

While the momentum operator $P_3^{(1+0)}$ creates/annihilates a single quantum energy $\hbar \omega$, the dilation operator $D_3^{1+0}$ and the special conformal operator $\mathfrak{K}_{{3}}^{(1+0)}$ create/annihilate up to two and three quanta of $2\hbar\omega$ and $3\hbar\omega$, respectively. 
Using Eqs. (\ref{creation_anni_P})-(\ref{creation_anni_KK}), we indeed find that 
the conformal algebra summarized in Table~\ref{tabeIFD10} is consistent with the creation/annihilation algebra prescribed in a one-dimensional simple harmonic oscillator. This leads us to consider the Fock space of the simple harmonic oscillator representation and discuss the coherent states of the simple harmonic oscillator in relation to the generators of the conformal algebra. 

We find that under translation $P^{(0+1)}_{3}$ the creation and annihilation operators transform as
\begin{align}
    a \rightarrow a^{\prime}_{P^{(1+0)}_{3}}=e^{ic P_{3}}ae^{-ic P_{3}}&=a +\frac{c}{\sqrt{2}},\\
    a^{\dagger} \rightarrow a^{\dagger\prime}_{P^{(1+0)}_{3}}=e^{ic P_{3}}a^{\dagger}e^{-ic P_{3}}&=a^{\dagger} + \frac{c}{\sqrt{2}},
\end{align}
where $c$ is a parameter representing the displacement. The displaced ground state of the SHO is known as the coherent state, as it is the eigenstate of the annihilation operator \cite{Glauber1963}. The new coherent vacuum $\ket{\Omega_{P_{3}}}$ by created by $a^{\dagger\prime}_{P^{(1+0)}_{3}}$ is given by
\begin{align}
    \ket{\Omega_{P_{3}}}&=e^{ic P_{3}}\ket{0}\\
    &=e^{-\frac{c^2}{4}}e^{-\frac{c}{\sqrt{2}}a^{\dagger}}\ket{0},
\end{align}
where $a\ket{0}=0$. The inner product of the trivial vacuum and the coherent vacuum is not zero as $\braket{0|\Omega_{P_{3}}} = 
e^{-\frac{c^2}{4}}$. Normalization of $\ket{\Omega_{P_{3}}}$ reads 
\begin{align}
\braket{\Omega_{P_{3}}|\Omega_{P_{3}}} &=1.
\end{align}
This normalization is consistent with the expectation that the normalization of the ground state $\braket{0|0} =1$ should be intact under the translation of the state.

We note that there is another type of coherent state besides Glauber's coherent state under the dilation $D^{(1+0)}_{3}$ operator.
It turns out that the dilation $D^{(1+0)}_{3}$ operator generates  the transformation of the creation and annihilation operators known as the  Bogoliubov-Valatin transformation \cite{umezawa1982thermo}, given by
\begin{align}
    a \rightarrow a^{\prime}_{D^{(1+0)}_{3}}=e^{i\alpha D_{3}}ae^{-i\alpha D_{3}}&=a\cosh{\alpha}-a^{\dagger}\sinh{\alpha}\label{aD1+0}\\
    a^{\dagger} \rightarrow a^{\dagger\prime}_{D^{(1+0)}_{3}}=e^{i\alpha D_{3}}a^{\dagger}e^{-i\alpha D_{3}}&=a^{\dagger}\cosh{\alpha}-a\sinh{\alpha},\label{adaggerD1+0}
\end{align}
where $\alpha$ is a parameter of dilatation. The new vacuum $\ket{\Omega_{D_{3}}}$ created by $a^{\dagger\prime}_{D^{(1+0)}_{3}}$ is given by
\begin{align}
    \ket{\Omega_{D_{3}}}&=e^{i\alpha D_{3}}\ket{0}\\
    &=\frac{e^{\alpha/2}}{\sqrt{\cosh\alpha}}\;
     e^{\left(\frac{\tanh\alpha}{2}\,a^{\dagger2}\right)}\ket{0},\label{vacuumD3}
\end{align}
and the normalization of $\ket{\Omega_{D_{3}}}$ reads
\begin{align}
    \braket{\Omega_{D_{3}}|\Omega_{D_{3}}}=&e^{\alpha}~.\label{innerproductvacuumD3}
\end{align}
The inner product of the dilated vacuum with the displaced vacuum is non-zero, as
\begin{align}   \braket{\Omega_{P_{3}}|\Omega_{D_{3}}} =&  \frac{e^{\alpha/2}}{\sqrt{\cosh\alpha}}\; e^{-\frac{c^2}{4}(1-\tanh\alpha)},\label{innerproductvacuumP3D3}
\end{align}
with normalization of vacuum $\braket{0|0}=1$, 
which shows that the two coherent states $\ket{\Omega_{P_{3}}}$ and $\ket{\Omega_{D_{3}}}$ are not orthogonal to each other as expected from their relationship with the ground state $\ket{0}$, namely, both coherent states are stemmed from the same ground state $\ket{0}$. We also note the consistency of Eq.\eqref{innerproductvacuumP3D3} with the expectation in the limit $\alpha \to 0$ or $c \to 0$ as  the state $\Omega_{D_3} \to \ket{0}$ or $\Omega_{P_3} \to \ket{0}$, respectively, in the corresponding limit. The derivation of Eq.\eqref{vacuumD3}, Eq.\eqref{innerproductvacuumD3} and Eq.\eqref{innerproductvacuumP3D3} are given in Appendix \ref{Dilated-vacuum}.

Under the special conformal transformation $\mathfrak{K}^{(1+0)}_{{3}}$, one may also find the corresponding coherent state which is not expected to be orthogonal to neither $\ket{\Omega_{P_{3}}}$ nor $\ket{\Omega_{D_{3}}}$ as we find that $\ket{\Omega_{\mathfrak{K}_{3}}}$ involves the exponent of up to the third order of $a$ and $a^\dagger$ while $\ket{\Omega_{P_{3}}}$ and $\ket{\Omega_{D_{3}}}$ involve the exponent of the first and second orders of $a$ and $a^\dagger$, respectively. We will defer further exploration of the coherent state $\ket{\Omega_{\mathfrak{K}_{3}}}$ in our future work.

\section{Interpolating Conformal Algebra in \texorpdfstring{$(1+1)$}{Lg} Dimensions}
\label{sec:conformal}

In contrast to the position operator discussed in the previous section, Sec.~\ref{sec:conformal0110}, the time itself cannot be implemented as an operator in quantum mechanics, according to Pauli's theorem \cite{Pauli1980}. Thus, in the $(1+1)$ dimensional relativistic theories, it is natural to demote the position operator from the operator level to the parameter level, as the time and position can mix at the same level due to the relativity. In Ref. \cite{Fubini1976}, indeed, the time itself was not an operator but a parameter, and the time-dependent field operators were introduced to quantify the field operators at equal time. It leads to the second quantization, i.e., the quantization of field operators, in contrast to the usual first quantization in quantum mechanics. This feature of the position parameter, in contrast to the position operator, leads to discussing the conformal algebra in the $(1+1)$ dimensional relativistic quantum field theories with the second quantization rather than the quantum mechanical quantization of the position operator itself in terms of the creation and annihilation operators as discussed in the previous section (Sec. \ref{sec:conformal0110}) for the $(1+0)$ dimensional conformal algebra. 

Based on the discussion of the lowest-dimensional conformal algebra in $(0+1)$ or $(1+0)$ dimension, we now consider the conformal algebra in the next higher dimension, i.e., the $(1+1)$ dimensions. Here, the generators are given by
\begin{align}
    P^{(1+1)}_{0}=&i\partial_{0},\\
    \mathfrak{K}^{(1+1)}_{{0}}=&i(x_{0}x_{0}\partial_{{0}}-2x_{3}x_{0}\partial_{3}+x_{3}x_{3}\partial_{0}),\\
    D^{(1+1)}=&ix_{0}\partial_{0}-ix_{3}\partial_{3},\\
    P^{(1+1)}_{3}=&i\partial_{3},\\
    \mathfrak{K}^{(1+1)}_{{3}}=&-i(x_{3}x_{3}\partial_{{3}}-2x_{0}x_{3}\partial_{0}+x_{0}x_{0}\partial_{3}),\\
    K_{3}^{(1+1)}=&-i(x_{3}\partial_{0}-x_{0}\partial_{3}).
\end{align}
Some of these generators can be given by 
combinations of the previously defined (0+1) and (1+0) dimensional generators in Sec.\ref{sec:conformal0110}, namely
\begin{align}
    P^{(1+1)}_{0}=&P^{(0+1)}_{0},\\
    D^{(1+1)}=&D^{(0+1)}_{0}+D^{(1+0)}_{3},\\
    P^{(1+1)}_{3}=&P^{(1+0)}_{3}.
\end{align}
As the $(1+1)$ dimensional dilatation generator $D^{(1+1)}$ combines the (0+1) and (1+0) dimensional dilatation generators, $D^{(0+1)}_{0}$ and $D^{(1+0)}_{3}$, and appears as a single generator, the total number of generators in $(1+1)$ dimensions gets reduced by one. 
However, the new generator known as the boost generator appears in $(1+1)$ dimensions, compensating the reduced number of generators by mixing the $(0+1)$ and $(1+0)$ dimensions as given by
\begin{align}
K_{3}^{(1+1)}=&-i(x_{3}\partial_{0}-x_{0}\partial_{3}).
\end{align}
Thus, the total number of generators in $(1+1)$ dimension is preserved as the sum of the number of generators in (0+1) and (1+0) dimensions, i.e. the total six (6=3+3) generators including the two special conformal generators $\mathfrak{K}^{(1+1)}_{{0}}$ and $\mathfrak{K}^{(1+1)}_{{3}}$ in $(1+1)$ dimension. We note here that 
the special conformal transformations $\mathfrak{K}^{(1+1)}_{{0}}$
and $\mathfrak{K}^{(1+1)}_{{3}}$ in $(1+1)$ dimension involves this boost generator as given by
\begin{align}
\mathfrak{K}^{(1+1)}_{{0}}=&\mathfrak{K}^{(0+1)}_{{0}}-x_{3}K_{3}^{(1+1)}-x_{3}x_{0}P^{(1+0)}_{3},\\
\mathfrak{K}^{(1+1)}_{{3}}=&\mathfrak{K}^{(1+0)}_{{3}}-x_{0}K_{3}^{(1+1)}+x_{3}x_{0}P^{(0+1)}_{0},
\end{align}
where the symmetry between time and space 
dimensions are manifest.
Computing the commutation relations between the new generator $K_{3}^{(1+1)}$ and each of the $(0+1)$ and $(1+0)$ generators, we find
\begin{align}
    \left[K_{3}^{(1+1)}, P^{(0+1)}_{0}\right] &= -iP^{(1+0)}_{3},\\
    \left[K_{3}^{(1+1)}, P^{(1+0)}_{3}\right] &= -iP^{(0+1)}_{0},\\
    \left[K_{3}^{(1+1)}, D^{(0+1)}_{0}\right] &=(x_{3}\partial_{0}+x_{0}\partial_{3}),\\
    \left[K_{3}^{(1+1)}, D^{(1+0)}_{3}\right] &=-(x_{3}\partial_{0}+x_{0}\partial_{3}),\\
    \left[K_{3}^{(1+1)}, \mathfrak{K}^{(0+1)}_{0}\right] &=(2x_{3}x_{0}\partial_{0}+x_{0}x_{0}\partial_{3}),\\
    \left[K_{3}^{(1+1)}, \mathfrak{K}^{(1+0)}_{3}\right] &=-(2x_{3}x_{0}\partial_{3}+x_{3}x_{3}\partial_{0}).
\end{align}
Using these commutation relations involving $K_{3}^{(1+1)}$ with the $(0+1)$ and $(1+0)$ generators, we derive the conformal algebras among the six generators in $(1+1)$ dimensions as shown in Table.~\ref{tabelinterpolationifd}, where we removed the cumbersome superscript $(1+1)$ in each of the six generators for simplicity. 

\begin{table}[h!]
\centering
\setlength{\tabcolsep}{0pt} 
\caption{\label{tabelinterpolationifd}
$1+1$ conformal algebra. Note here that the cumbersome superscript $(1+1)$ is removed for all generators.}
\begin{tabular}{ |>{\centering\arraybackslash}p{1cm}||>{\centering\arraybackslash}p{1cm}|>{\centering\arraybackslash}p{1cm}|>{\centering\arraybackslash}p{1cm}|>{\centering\arraybackslash}p{1cm}|>{\centering\arraybackslash}p{1cm}|>{\centering\arraybackslash}p{1cm}| } 
 \hline
\rule{0pt}{16pt} & $\mathfrak{K}_{{0}}$   & $P_{{0}}$ &  $D$& $\mathfrak{K}_{{3}}$ & $P_{{3}}$& $K_{{3}}$\\
 \hline
  \hline
 \rule{0pt}{16pt}$\mathfrak{K}_{{0}}$ & \cellcolor{blue!20}0&\cellcolor{blue!20}$-2iD$&\cellcolor{blue!20}$-i\mathfrak{K}_{{0}}$&\cellcolor{red!20}0&\cellcolor{red!20}$-2iK_{{3}}$&\cellcolor{red!20}${i\mathfrak{K}_{{3}}}$\\
  \hline 
 \rule{0pt}{16pt}  $P_{{0}}$ &\cellcolor{blue!20}$2iD$&\cellcolor{blue!20}0&\cellcolor{blue!20}$iP_{{0}}$&\cellcolor{red!20}${-2iK_{{3}}}$&\cellcolor{red!20}0&\cellcolor{red!20}${iP_{{3}}}$\\
 \hline
 \rule{0pt}{16pt}$D$ &\cellcolor{blue!20}$i\mathfrak{K}_{{0}}$&\cellcolor{blue!20}$-iP_{{0}}$&\cellcolor{blue!20}0&\cellcolor{red!20}$i\mathfrak{K}_{{3}}$&\cellcolor{red!20}$-iP_{{3}}$&\cellcolor{red!20}0\\
 \hline
 \rule{0pt}{16pt}$\mathfrak{K}_{{3}}$ &\cellcolor{red!20}0&\cellcolor{red!20}${2iK_{{3}}}$&\cellcolor{red!20}$-i\mathfrak{K}_{{3}}$&\cellcolor{cyan!20}0&\cellcolor{cyan!20}$2iD$&\cellcolor{cyan!20}${i\mathfrak{K}_{{0}}}$\\
 \hline 
 \rule{0pt}{16pt}$P_{{3}}$ &\cellcolor{red!20}$2iK_{{3}}$&\cellcolor{red!20}0&\cellcolor{red!20}$iP_{{3}}$&\cellcolor{cyan!20}$-2iD$&\cellcolor{cyan!20}0&\cellcolor{cyan!20}${iP_{{0}}}$\\
 \hline 
 \rule{0pt}{16pt}$K_{{3}}$ &\cellcolor{red!20}${-i\mathfrak{K}_{{3}}}$&\cellcolor{red!20}${-iP_{{3}}}$&\cellcolor{red!20}0&\cellcolor{cyan!20}${-i\mathfrak{K}_{{0}}}$&\cellcolor{cyan!20}${-iP_{{0}}}$&\cellcolor{cyan!20}0\\
 \hline 
\end{tabular}
\end{table}

In Table.~\ref{tabelinterpolationifd}, we note the apparent symmetry highlighted by colors in ``blue/cyan" and ``red" denoting the first/second block diagonal part and the two block off-diagonal parts. The first block diagonal part marked by ``blue" coincides with the (0+1) conformal algebra in Table.~\ref{tabeIFD01}, where the generators $\mathfrak{K}_{{0}}$, $P_0$ and $D$ in Table.~\ref{tabelinterpolationifd} correspond to the respective (0+1) conformal generators $\mathfrak{K}^{(0+1)}_{{0}}$, $P^{(0+1)}_{0}$ and $D^{(0+1)}_{0}$. Likewise, the second block diagonal part marked by ``cyan" appears the same as the first block diagonal part with the respective replacement of $\mathfrak{K}_{{0}}$, $P_0$, and $D$ by 
$\mathfrak{K}_{{3}}$, $P_3$ and $K_3$ and the correspondence between $D$ and $K_3$. Similarly, the two block off-diagonal parts, marked in red, correspond to each other, with an apparent symmetry between them.


For the relativistic quantum field theoretic description of the conformal algebra, we may first briefly summarize the two distinguished forms of the relativistic dynamics proposed by Dirac  
in 1949 \cite{Dirac1949}, i.e. the instant form ($x^{0}=0$) and the front form ($x^{+}=(x^{0}+x^{3})/\sqrt{2}$=0). 
While the instant form dynamics (IFD) of quantum field theory is quantized at the equal time $t=x^{0}$, the light-front dynamics (LFD) is quantized at the equal light-front time $\tau \equiv (x^{0}+x^{3})/\sqrt{2}=x^{+}$ ($c=1$ unit is taken here).

The LFD is distinguished from the IFD due to the drastic difference in the energy-momentum dispersion relation. For a particle of mass $m$ in $(1+1)$ dimensions, the energy-momentum relation at equal-$\tau$ (LFD) is rational as given by
\begin{align}
  k^{-}=\dfrac{m^{2}}{k^{+}}, \label{eqn:E-P_relation_LF}
\end{align}
where the light-front energy $k^{-}=(k^{0}-k^{3})/\sqrt{2}$ is conjugate to $\tau$, and the light-front momentum $k^{+}=(k^{0}+k^{3})/\sqrt{2}$,
while the corresponding energy-momentum dispersion relation of the particle at equal-$t$ (IFD) is irrational, as given by
\begin{align}
  k^{0}=\sqrt{k_z^{2}+m^{2}}, \label{eqn:E-P_relation_IF}
\end{align}
where the energy $k^{0}$ is conjugate to $t$ and the longitudinal momentum $k_z = k^{3}$. 
The remarkable feature of the LFD energy-momentum relation is the correlation between the signs of $k^{-}$ and $k^{+}$. Namely, for the positive light-front energy $k^{-}$, the light-front longitudinal momentum $k^{+}$ must be non-negative.
Except the light-front zero-modes $k^{+}=0$, the light-front longitudinal momentum must be positive $k^{+}>0$. This feature makes the LFD quite distinct from the typical IFD, which allows both positive and negative longitudinal momentum $-\infty < k^3 < \infty$. The sign correlation of $k^+$ and $k^-$ in LFD yields dramatic consequences in the relativistic quantum field theoretic vacuum characteristics as discussed in various literature~ \cite{Brodsky_1998, BrodskyLightFrontMethodsandNonPerturbativeQCD, harindranath1998introductionlightfrontdynamicspedestrians, ji2023relativistic}. 
Even the structure of the Poincar\'e algebra is drastically changed in the LFD compared to the IFD, which has been discussed extensively in the past~ \cite{Ji2001}.
LFD saves tremendous dynamic efforts in solving the relativistic quantum field theoretic problems by maximizing the number of kinematic generators. 
A paramount example may be found in solving the mass gap equation in the $(1+1)$ dimensional QCD at a large number of color degrees of freedom $N_c \to \infty$, which is known as the 'tHooft model~ \cite{THOOFT1974461, Ji2021QCD, ji2023relativistic}. Effectively, the LFD maximizes the effectiveness of QCD description in the spectroscopy and structure studies of hadrons, which reflects the full Poincar\'e symmetries.

In the LFD, the conformal generators in $(1+1)$ dimensions can be identified as 
$P_{\pm}=\frac{P_{0}\pm P_{3}}{\sqrt{2}}$, $\mathfrak{K}_{\pm}=\frac{\mathfrak{K}_{0}\mp \mathfrak{K}_{3}}{\sqrt{2}}$ and $D_{\pm}=\frac{D\pm{K_{3}}}{\sqrt{2}}$ and one can find the commutation relation among themselves 
as given in Table~\ref{tabelinterpolationlfd}. 
We note here that the generators, $P_{\pm}$ and $\mathfrak{K}_{\mp} $, translate the light-front coordinates $x^{\pm}$ and $\frac{1}{x^{\pm}}$, respectively, as one may illustrate from the M\"{o}bius transformation in LFD. The dilatation generator $D_{\pm}$ is correspondingly fixed to keep the symmetry between $+$ and $-$ spaces in LFD.

\begin{table}[h!]
\centering
\setlength{\tabcolsep}{0pt} 
\caption{\label{tabelinterpolationlfd}$1+1$ conformal algebra in LFD.}
\scalebox{0.9}{
\begin{tabular}{ |>{\centering\arraybackslash}p{1cm}||>{\centering\arraybackslash}p{1.3cm}|>{\centering\arraybackslash}p{1.5cm}|>{\centering\arraybackslash}p{1.3cm}|>{\centering\arraybackslash}p{1.3cm}|>{\centering\arraybackslash}p{1.5cm}|>{\centering\arraybackslash}p{1.3cm}| } 
 \hline
 \rule{0pt}{16pt} & $\mathfrak{K}_{{+}}$ & $P_{{+}}$ & $D_{+}$ & $\mathfrak{K}_{{-}}$ & $P_{{-}}$ & $D_{{-}}$ \\
 \hline
 \hline
 \rule{0pt}{16pt}$\mathfrak{K}_{{+}}$ & \cellcolor{blue!20}0 & \cellcolor{blue!20}$-2\sqrt{2}iD_{+}$ & \cellcolor{blue!20}$-\sqrt{2}i\mathfrak{K}_{{+}}$ & \cellcolor{red!20}0 & \cellcolor{red!20}0 & \cellcolor{red!20}0\\
 \hline
 \rule{0pt}{16pt}$P_{{+}}$ & \cellcolor{blue!20}$2\sqrt{2}iD_{+}$ & \cellcolor{blue!20}0 & \cellcolor{blue!20}$\sqrt{2}iP_{{+}}$ & \cellcolor{red!20}0 & \cellcolor{red!20}0 & \cellcolor{red!20}0\\
 \hline
 \rule{0pt}{16pt}$D_{+}$ & \cellcolor{blue!20}$\sqrt{2}i\mathfrak{K}_{{+}}$ & \cellcolor{blue!20}$-\sqrt{2}iP_{{+}}$ & \cellcolor{blue!20}0 & \cellcolor{red!20}0 & \cellcolor{red!20}0 & \cellcolor{red!20}0\\
 \hline
 \rule{0pt}{16pt}$\mathfrak{K}_{{-}}$ & \cellcolor{red!20}0 & \cellcolor{red!20}0 & \cellcolor{red!20}0 & \cellcolor{cyan!20}0 & \cellcolor{cyan!20}$-2\sqrt{2}iD_{-}$ & \cellcolor{cyan!20}$-\sqrt{2}i\mathfrak{K}_{{-}}$\\
 \hline
 \rule{0pt}{16pt}$P_{{-}}$ & \cellcolor{red!20}0 & \cellcolor{red!20}0 & \cellcolor{red!20}0 & \cellcolor{cyan!20}$2\sqrt{2}iD_{-}$ & \cellcolor{cyan!20}0 & \cellcolor{cyan!20}$\sqrt{2}iP_{{-}}$\\
 \hline
 \rule{0pt}{16pt}$D_{-}$ & \cellcolor{red!20}0 & \cellcolor{red!20}0 & \cellcolor{red!20}0  & \cellcolor{cyan!20}$\sqrt{2}i\mathfrak{K}_{{-}}$& \cellcolor{cyan!20}$-\sqrt{2}iP_{{-}}$ & \cellcolor{cyan!20}0\\
 \hline
\end{tabular}}
\end{table}
From the perspectives of relativistic quantum invariance, it would be useful to find the correspondence between the IFD and the LFD 
for the (1+1)d conformal algebra exhibited in Tables ~\ref{tabelinterpolationifd} and ~\ref{tabelinterpolationlfd}, respectively. 

In this work, we interpolate the $(1+1)$dimensional conformal algebra between IFD and LFD. As one finds a distinguished feature in each form of the dynamics in analyzing the Poincaré algebra with respect to the number of kinematic generators, one may further see the characteristics of the conformal generators, such as the dilatation and the special conformal transformations, in each form of the dynamics from the investigation of the interpolating conformal algebra between IFD and LFD.

To interpolate the forms of relativistic quantum field theory between IFD and LFD, we take the following interpolating spacetime coordinates  \cite{Ji2001, Hornbostel1992, Ji1996, Ji2012, Ji2015EM, Ji2015SP, Ji2018QED, Ji2021QCD} which is defined as a transformation from the ordinary spacetime coordinates $x^{\muT}=\mathcal{R}^{\muT}_{\phantom{\mu}{\nu}}x^{\nu}$, i.e.,
\begin{align}\label{eqn:interpolation_angle_definition}
  \begin{pmatrix}
    x^{\hat{+}}\\
    x^{\hat{-}}
  \end{pmatrix}=
  \begin{pmatrix}
    \cos\delta  & \sin\delta \\
    \sin\delta  & -\cos\delta
  \end{pmatrix}
  \begin{pmatrix}
    x^{0}\\
    x^{3}
  \end{pmatrix},
\end{align}
in which the interpolation angle is allowed to run from $0^\circ$ (IFD) through $45^\circ$ (LFD), $0\le \delta \le \frac{\pi}{4}$. The interpolating coordinates $x^{\itP{\pm}}$ in the limit $\delta\rightarrow\pi/4$ become the light-front coordinates $x^{\pm}=(x^{0}\pm x^{3})/\sqrt{2}$ without  ``\textasciicircum''. Note that we interpolate from $-x^3$, so in the limit $\delta\rightarrow\pi/4$, the interpolating coordinates $x^{\itP{-}}$ is $-x^3$ (or $-z$) axis. 

 In this interpolating basis, the metric becomes
\begin{align}\label{eqn:g_munu_interpolation}
  g^{\hat{\mu}\hat{\nu}}
  = g_{\hat{\mu}\hat{\nu}}
  =
  \begin{pmatrix}
    \mathbb{C}  & \mathbb{S} \\
    \mathbb{S}  & -\mathbb{C}
  \end{pmatrix},
\end{align}
where $\mathbb{S}=\sin2\delta$ and $\mathbb{C}=\cos2\delta$. The lower-index variables $x_{\hat{+}}$ and $x_{\hat{-}}$ are related to the upper-index variables as $x_{\hat{+}}=g_{\hat{+}\hat{+}}x^{\hat{+}}+g_{\hat{+}\hat{-}}x^{\hat{-}}=\mathbb{C}x^{\hat{+}}+\mathbb{S}x^{\hat{-}}$ and $x_{\hat{-}}=g_{\hat{-}\hat{+}}x^{\hat{+}}+g_{\hat{-}\hat{-}}x^{\hat{-}}=-\mathbb{C}x^{\hat{-}}+\mathbb{S}x^{\hat{+}}$.
The details of the relationship between the
interpolating variables and the usual spacetime variables
in (3+1)d can be seen in our previous works  \cite{Ji2001, Hornbostel1992, Ji1996, Ji2012, Ji2015EM, Ji2015SP, Ji2018QED, Ji2021QCD} although we focus on the (1+1)d conformal algebra in this work.

To make the conformal algebra into a simpler form in (1+1)d, we define the following generators in four-dimensional projective spacetime of IFD:
\begin{align}\label{Jab}
  J_{ab}&=
  \begin{pmatrix}
  0&-D&\frac{-\mathfrak{K}_0}{L\sqrt{2}}&\frac{-\mathfrak{K}_3}{L\sqrt{2}}\\
  D&0&\frac{LP_0}{\sqrt{2}}&\frac{LP_3}{\sqrt{2}}\\
    \frac{\mathfrak{K}_0}{L\sqrt{2}}&\frac{-LP_0}{\sqrt{2}}&0  & K_{3}\\
    \frac{\mathfrak{K}_3}{L\sqrt{2}}&\frac{-LP_3}{\sqrt{2}}&-K_{3} & 0
  \end{pmatrix}_{4\times4},
\end{align}
where $L$ is any constant with the dimension of length~ \cite{Fubini1976} which we take $L=1$ in this work, $J_{ab}=-J_{ba}$ (i.e. antisymmetric matrix) and $a,b\in\{-2,-1,0,3\}$. 
In (3+1)d, the six-dimensional projective spacetime representations in IFD were discussed in Refs. \cite{SalamMack1969, Weinberg_2010}.
All $J_{ab}$ elements are conformal generators in $(1+1)d$. We omit here any cumbersome $(1+1)$ superscript for notational simplicity, without which $J_{ab} = J^{(1+1)}_{ab}$, as the generators here are obviously in $(1+1)d$. The generators $J_{ab}$ obey the $SO(2,1+1)$ commutation relations:
  \begin{align}\label{JabalgebraIFD}
      \left[J_{{a}{b}},J_{{c}{d}}\right]=-i\left(g_{{b}{d}}J_{{a}{c}}-g_{{b}{c}}J_{{a}{d}}+g_{{a}{c}}J_{{b}{d}}-g_{{a}{d}}J_{{b}{c}}\right)
  \end{align}
where, 
  \begin{align}\label{metric}
      g_{ab}=\begin{pmatrix}
  0&-1&0&0\\
  -1&0&0&0\\
  0&0&1&0\\
  0&0&0&-1\\
  \end{pmatrix}_{4\times4}.
  \end{align}
A similar simplification of conformal algebra with a fully diagonal metric was mentioned in various literature sources \cite{SalamMack1969, Francesco}. We may split the $4\times4$ matrix representation of $J_{ab}$ ($a,b\in\{-2,-1,0,3\}$) into the two separate $3\times3$ matrix representations of $J^{(0)}_{pq}$ ($p,q\in\{1,2,3\}$) and $J^{(3)}_{rs}$ ($r,s\in\{1,2,3\}$) in which the respective superscripts (0) and (3) signify the involvement of time and space translations in (1+1)d, corresponding the matrix elements of $J^{(0)}_{pq}$ ($p,q\in\{1,2,3\}$) and $J^{(3)}_{rs}$ ($r,s\in\{1,2,3\}$) with the matrix elements of $J_{ab}$ ($a,b\in\{-2,-1,0,3\}$) as follows:
\begin{align}
\label{correspondence-of-matrix-elements}
    J^{(0)}_{12} =&J_{-2-1},&& J^{(3)}_{12} = -J_{03}, \nonumber\\
    J^{(0)}_{23} =&J_{-10} ,&& J^{(3)}_{23}=J_{-13},\nonumber\\
    J^{(0)}_{31}=&J_{0-2} ,&& J^{(3)}_{31}=J_{-23}.
\end{align}
Note here that both $J^{(0)}_{pq}$ and $J^{(3)}_{rs}$ are antisymmetric matrices, i.e. $J^{(0)}_{pq} = - J^{(0)}_{qp}$ and $J^{(3)}_{rs}=-J^{(3)}_{sr}$, inheriting the characteristic antisymmetric property of $J_{ab}$, i.e. $J_{ab}=-J_{ba}$.  
This splitting of time and space 
translations with the two $3\times3$ matrices denoted by $J^{(0)}_{pq}$ and $J^{(3)}_{pq}$, respectively, allow us the interpolation of the conformal algebra in (1+1)d as we have interpolated the time and space translations, $J^{(0)}_{23}=\frac{P_0}{\sqrt{2}}$ and $J^{(3)}=\frac{P_3}{\sqrt{2}}$. Indeed, it is remarkable to note that the arrangement of each and every element of $J^{(0)}_{pq}$ and $J^{(3)}_{pq}$ is characteristically consistent. In particular, not only the correspondence between $J^{(0)}_{13}=\frac{-\mathfrak{K}_0}{\sqrt{2}}$ and $J^{(3)}=\frac{\mathfrak{K}_3}{\sqrt{2}}$ matches consistently with the correspondence between $J^{(0)}_{23}=\frac{P_0}{\sqrt{2}}$ and $J^{(3)}=\frac{P_3}{\sqrt{2}}$ but also the correspondence between $J^{(0)}_{12}=-D$ and $J^{(3)}_{12}=K^3=-K_3$ provides the consistent characteristics of the physical length contraction and time dilation physical phenomena realized by the dilation of space and time. 
We may summarize these characteristic correspondence between $J^{(0)}_{pq}$ and $J^{(3)}_{pq}$ with the explicit representations of $J^{(0)}_{pq}$ and $J^{(3)}_{pq}$ as $3\times3$ matrices given by 
\begin{align}
  J^{(0)}_{p q}=
  \begin{pmatrix}
  0&-D&\frac{-\mathfrak{K}_0}{\sqrt{2}}\\
  D&0&\frac{P_0}{\sqrt{2}}\\
    \frac{\mathfrak{K}_0}{\sqrt{2}}&\frac{-P_0}{\sqrt{2}}&0  
  \end{pmatrix}_{3\times3}; J^{(3)}_{pq}=
  \begin{pmatrix}
  0&-K_3&\frac{\mathfrak{K}_3}{\sqrt{2}}\\
  K_3&0&\frac{P_3}{\sqrt{2}}\\
    \frac{-\mathfrak{K}_3}{\sqrt{2}}&\frac{-P_3}{\sqrt{2}}&0  
  \end{pmatrix}_{3\times3}.
\end{align}
We note here that $J^{(0)}_{p q}$ is just the $3\times3$ block of the $4\times4$ matrix $J_{ab}$ (see Eq.\eqref{Jab}), reflecting the correspondence of matrix elements given by Eq.\eqref{correspondence-of-matrix-elements}.   
Introducing the metric for the $3\times3$ block of $J^{(0)}_{p q}$ as well as $J^{(3)}_{p q}$ consistently given by   
\begin{align}
      g_{pq}=\begin{pmatrix}
  0&-1&0\\
  -1&0&0\\
  0&0&1\\
  \end{pmatrix}_{3\times3},
  \end{align}
We find the identical structures of the (1+1)d conformal algebra
as summarized below:
\begin{align}
      \left[J^{(0)}_{{p}{q}},J^{(0)}_{{r}{s}}\right]&=-i\left(g_{{q}{s}}J^{(0)}_{{p}{r}}-g_{{q}{r}}J^{(0)}_{{p}{s}}+g_{{p}{r}}J^{(0)}_{{q}{s}}-g_{{p}{s}}J^{(0)}_{{q}{r}}\right),\label{algebraJpq1}\\
      \left[J^{(3)}_{{p}{q}},J^{(3)}_{{r}{s}}\right]&=-i\left(g_{{q}{s}}J^{(0)}_{{p}{r}}-g_{{q}{r}}J^{(0)}_{{p}{s}}+g_{{p}{r}}J^{(0)}_{{q}{s}}-g_{{p}{s}}J^{(0)}_{{q}{r}}\right),\label{algebraJpq2}\\
      \left[J^{(0)}_{{p}{q}},J^{(3)}_{{r}{s}}\right]&=-i\left(g_{{q}{s}}J^{(3)}_{{p}{r}}-g_{{q}{r}}J^{(3)}_{{p}{s}}+g_{{p}{r}}J^{(3)}_{{q}{s}}-g_{{p}{s}}J^{(3)}_{{q}{r}}\right),\label{algebraJpq3}\\
      \left[J^{(3)}_{{p}{q}},J^{(0)}_{{r}{s}}\right]&=-i\left(g_{{q}{s}}J^{(3)}_{{p}{r}}-g_{{q}{r}}J^{(3)}_{{p}{s}}+g_{{p}{r}}J^{(3)}_{{q}{s}}-g_{{p}{s}}J^{(3)}_{{q}{r}}\right)\label{algebraJpq4}.
  \end{align}
This result confirms the validity of Eq.\eqref{correspondence-of-matrix-elements} which corresponds the matrix elements of $3\times3$ matrices
 $J^{(0)}_{pq}$ ($p,q\in\{1,2,3\}$) and $J^{(3)}_{rs}$ ($r,s\in\{1,2,3\}$) with the $4\times4$ matrix $J_{ab}$ ($a,b\in\{-2,-1,0,3\}$).

With this preparation of the time and space conformal algebraic $3\times3$ matrices in IFD, $J^{(0)}_{pq}$ ($p,q\in\{1,2,3\}$) and $J^{(3)}_{rs}$ ($r,s\in\{1,2,3\}$), we now interpolate them as we have done in the Poincar\'e algebra previously~ \cite{Ji2001, Ji2025}. The interpolation of the (1+1)d conformal algebra  between IFD and LFD is then given by  
\begin{align}
    J^{(\hat{+})}_{pq}&=J^{(0)}_{pq}\cos{\delta}+J^{(3)}_{pq}\sin{\delta}\\
    J^{(\hat{-})}_{pq}&=J^{(0)}_{pq}\sin{\delta}-J^{(3)}_{pq}\cos{\delta}.
\end{align}
which yields
\begin{align}
    J^{(\hat{\pm})}_{pq}=
  \begin{pmatrix}
  0&-D_{\hat{\pm}}&\frac{-\mathfrak{K}_{\hat{\pm}}}{\sqrt{2}}\\
  D_{\hat{\pm}}&0&\frac{P_{\hat{\pm}}}{\sqrt{2}}\\
    \frac{\mathfrak{K}_{\hat{\pm}}}{\sqrt{2}}&\frac{-P_{\hat{\pm}}}{\sqrt{2}}&0  
  \end{pmatrix}_{3\times3},
\end{align}
where, $P_{\hat{+}}=P_0\cos{\delta}+P_3\sin{\delta}$, $P_{\hat{-}}=P_0\sin{\delta}-P_3\cos{\delta}$, $\mathfrak{K}_{\hat{+}}=\mathfrak{K}_0\cos{\delta}-\mathfrak{K}_3\sin{\delta}$,  $\mathfrak{K}_{\hat{-}}=\mathfrak{K}_0\sin{\delta}+\mathfrak{K}_3\cos{\delta}$, $D_{\hat{+}}=D\cos\delta+K_{3}\sin\delta$, and $D_{\hat{-}}=D\sin\delta-K_{3}\cos\delta$. 

In the IFD limit $\delta\rightarrow0$ (or $\mathbb{S}\rightarrow0$ \& $\mathbb{C}\rightarrow1$), we have $P_{\hat{+}}\rightarrow P_{0}$, $P_{\hat{-}}\rightarrow -P_{3}$, $\mathfrak{K}_{\hat{+}}\rightarrow \mathfrak{K}_{0}$, $\mathfrak{K}_{\hat{-}}\rightarrow \mathfrak{K}_{3}$, $D_{\hat{+}}\rightarrow D$, and $D_{\hat{-}}\rightarrow -K_{3}$. Likewise, in the LFD limit $\delta\rightarrow\frac{\pi}{4}$ (or $\mathbb{S}\rightarrow1$ \& $\mathbb{C}\rightarrow0$), we have $P_{\hat{\pm}}\rightarrow P_{\pm}=\frac{P_0\pm P_3}{\sqrt{2}}$, $\mathfrak{K}_{\hat{\pm}}\rightarrow \mathfrak{K}_{\pm}=\frac{\mathfrak{K}_0\mp \mathfrak{K}_3}{\sqrt{2}}$, and $D_{\hat{\pm}}\rightarrow D_{\pm}=\frac{D \pm K_3}{\sqrt{2}}$. Then the $(1+1)$ conformal algebra in the interpolation between IFD and LFD reads,
\begin{widetext}
\begin{center}
\begin{align}
    \left[J^{\hat{+}}_{pq},J^{\hat{+}}_{rs}\right] &=-ic(2-\mathbb{C})\left(g_{{q}{s}}J^{\hat{+}}_{{p}{r}}-g_{{q}{r}}J^{\hat{+}}_{{p}{s}}+g_{{p}{r}}J^{\hat{+}}_{{q}{s}}-g_{{p}{s}}J^{\hat{+}}_{{q}{r}}\right)+is\mathbb{C}\left(g_{{q}{s}}J^{\hat{-}}_{{p}{r}}-g_{{q}{r}}J^{\hat{-}}_{{p}{s}}+g_{{p}{r}}J^{\hat{-}}_{{q}{s}}-g_{{p}{s}}J^{\hat{-}}_{{q}{r}}\right),\\
    \left[J^{\hat{-}}_{pq},J^{\hat{-}}_{rs}\right] &=-ic\mathbb{C}\left(g_{{q}{s}}J^{\hat{+}}_{{p}{r}}-g_{{q}{r}}J^{\hat{+}}_{{p}{s}}+g_{{p}{r}}J^{\hat{+}}_{{q}{s}}-g_{{p}{s}}J^{\hat{+}}_{{q}{r}}\right)-is(2+\mathbb{C})\left(g_{{q}{s}}J^{\hat{-}}_{{p}{r}}-g_{{q}{r}}J^{\hat{-}}_{{p}{s}}+g_{{p}{r}}J^{\hat{-}}_{{q}{s}}-g_{{p}{s}}J^{\hat{-}}_{{q}{r}}\right),\\
    \left[J^{\hat{+}}_{pq},J^{\hat{-}}_{rs}\right] &=is\mathbb{C}\left(g_{{q}{s}}J^{\hat{+}}_{{p}{r}}-g_{{q}{r}}J^{\hat{+}}_{{p}{s}}+g_{{p}{r}}J^{\hat{+}}_{{q}{s}}-g_{{p}{s}}J^{\hat{+}}_{{q}{r}}\right)+ic\mathbb{C}\left(g_{{q}{s}}J^{\hat{-}}_{{p}{r}}-g_{{q}{r}}J^{\hat{-}}_{{p}{s}}+g_{{p}{r}}J^{\hat{-}}_{{q}{s}}-g_{{p}{s}}J^{\hat{-}}_{{q}{r}}\right),\\
    \left[J^{\hat{-}}_{pq},J^{\hat{+}}_{rs}\right] &=is\mathbb{C}\left(g_{{q}{s}}J^{\hat{+}}_{{p}{r}}-g_{{q}{r}}J^{\hat{+}}_{{p}{s}}+g_{{p}{r}}J^{\hat{+}}_{{q}{s}}-g_{{p}{s}}J^{\hat{+}}_{{q}{r}}\right)+ic\mathbb{C}\left(g_{{q}{s}}J^{\hat{-}}_{{p}{r}}-g_{{q}{r}}J^{\hat{-}}_{{p}{s}}+g_{{p}{r}}J^{\hat{-}}_{{q}{s}}-g_{{p}{s}}J^{\hat{-}}_{{q}{r}}\right),
\end{align}
\end{center}
\end{widetext}

where $s=\sin{\delta}$ and $c=\cos{\delta}$. A comprehensive table of the fifteen commutation relations among the covariant components of the Conformal generators in interpolation form is presented below in the Table.~\ref{tabel1+1interpolationlfd}:

\begin{widetext}
\begin{center}
\begin{table}[h!]
\centering
\setlength{\tabcolsep}{0pt} 
\caption{\label{tabel1+1interpolationlfd}$1+1$ conformal algebra in the interpolation form}
\scalebox{0.64}{
\begin{tabular}{ |>{\centering\arraybackslash}p{1cm}||>{\centering\arraybackslash}p{4.4cm}|>{\centering\arraybackslash}p{4.5cm}|>{\centering\arraybackslash}p{4.4cm}|>{\centering\arraybackslash}p{4.4cm}|>{\centering\arraybackslash}p{4.5cm}|>{\centering\arraybackslash}p{4.4cm}| } 
\hline
 \rule{0pt}{25pt} & $\mathfrak{K}_{\hat{+}}$& $P_{\hat{+}}$& $D_{\hat{+}}$  & $\mathfrak{K}_{\hat{-}}$  &  $P_{\hat{-}}$ & $D_{\hat{-}}$ \\
 \hline
  \hline
 \rule{0pt}{25pt}$\mathfrak{K}_{\hat{+}}$  & \cellcolor{blue!20}$0$ & \cellcolor{blue!20}$-2i[(c+\mathbb{S}s)D_{\hat{+}}+(s-\mathbb{S}c)D_{\hat{-}}]$& \cellcolor{blue!20} $-i[(c+\mathbb{S}s)\mathfrak{K}_{\hat{+}}+(s-\mathbb{S}c)\mathfrak{K}_{\hat{-}}]$& \cellcolor{red!20}$0$ & \cellcolor{red!20}$2i\mathbb{C}[sD_{\hat{+}}-cD_{\hat{-}}]$ & \cellcolor{red!20} $i\mathbb{C}[s\mathfrak{K}_{\hat{+}}-c\mathfrak{K}_{\hat{-}}]$\\
 \hline 
 \rule{0pt}{25pt}  $P_{\hat{+}}$ & \cellcolor{blue!20} $2i[(c+\mathbb{S}s)D_{\hat{+}}+(s-\mathbb{S}c)D_{\hat{-}}]$& \cellcolor{blue!20} $0$& \cellcolor{blue!20} $i[(c+\mathbb{S}s)P_{\hat{+}}+(s-\mathbb{S}c)P_{\hat{-}}]$& \cellcolor{red!20} $-2i\mathbb{C}[sD_{\hat{+}}-cD_{\hat{-}}]$& \cellcolor{red!20} $0$ & \cellcolor{red!20} $-i\mathbb{C}[sP_{\hat{+}}-cP_{\hat{-}}]$\\
 \hline 
  \rule{0pt}{25pt}$D_{\hat{+}}$  & \cellcolor{blue!20}$i[(c+\mathbb{S}s)\mathfrak{K}_{\hat{+}}+(s-\mathbb{S}c)\mathfrak{K}_{\hat{-}}]$ & \cellcolor{blue!20}$-i[(c+\mathbb{S}s)P_{\hat{+}}+(s-\mathbb{S}c)P_{\hat{-}}]$& \cellcolor{blue!20} $0$& \cellcolor{red!20}$-i\mathbb{C}[s\mathfrak{K}_{\hat{+}}-c\mathfrak{K}_{\hat{-}}]$ & \cellcolor{red!20} $i\mathbb{C}[sP_{\hat{+}}-cP_{\hat{-}}]$& \cellcolor{red!20} $0$\\
 \hline 
 \rule{0pt}{25pt}$\mathfrak{K}_{\hat{-}}$ & \cellcolor{red!20} $0$& \cellcolor{red!20} $2i\mathbb{C}[sD_{\hat{+}}-cD_{\hat{-}}]$& \cellcolor{red!20}$i\mathbb{C}[s\mathfrak{K}_{\hat{+}}-c\mathfrak{K}_{\hat{-}}]$  & \cellcolor{cyan!20} $0$& \cellcolor{cyan!20}$-2i[(c-\mathbb{S}s)D_{\hat{+}}+(s+\mathbb{S}c)D_{\hat{-}}]$ & \cellcolor{cyan!20} $-i[(c-\mathbb{S}s)\mathfrak{K}_{\hat{+}}+(s+\mathbb{S}c)\mathfrak{K}_{\hat{-}}]$\\
 \hline 
 \rule{0pt}{25pt}$P_{\hat{-}}$  & \cellcolor{red!20} $-2i\mathbb{C}[sD_{\hat{+}}-cD_{\hat{-}}]$& \cellcolor{red!20}$0$& \cellcolor{red!20}$-i\mathbb{C}[sP_{\hat{+}}-cP_{\hat{-}}]$ & \cellcolor{cyan!20}$2i[(c-\mathbb{S}s)D_{\hat{+}}+(s+\mathbb{S}c)D_{\hat{-}}]$ & \cellcolor{cyan!20} $0$& \cellcolor{cyan!20} $i[(c-\mathbb{S}s)P_{\hat{+}}+(s+\mathbb{S}c)P_{\hat{-}}]$\\
 \hline
 \rule{0pt}{18pt}$D_{\hat{-}}$ & \cellcolor{red!20} $-i\mathbb{C}[s\mathfrak{K}_{\hat{+}}-c\mathfrak{K}_{\hat{-}}]$& \cellcolor{red!20} $i\mathbb{C}[sP_{\hat{+}}-cP_{\hat{-}}]$& \cellcolor{red!20} $0$& \cellcolor{cyan!20} $i[(c-\mathbb{S}s)\mathfrak{K}_{\hat{+}}+(s+\mathbb{S}c)\mathfrak{K}_{\hat{-}}]$& \cellcolor{cyan!20} $-i[(c-\mathbb{S}s)P_{\hat{+}}+(s+\mathbb{S}c)P_{\hat{-}}]$& \cellcolor{cyan!20} $0$\\
 \hline  
\end{tabular}}
\end{table}
\end{center}
\end{widetext}

In the limit $\delta\rightarrow0$; $\mathbb{C}\rightarrow0;~\mathbb{S}\rightarrow1$, we recover the commutation relations Eqs. \eqref{algebraJpq1}, \eqref{algebraJpq2}, \eqref{algebraJpq3}, \eqref{algebraJpq4} among all IFD conformal generators in two dimensions as given explicitly in the Table. \ref{tabelinterpolationifd}. In the limit $\delta\rightarrow\frac{\pi}{4}$; $\mathbb{C}\rightarrow1;~\mathbb{S}\rightarrow0$, we recover the commutation relations among all LFD conformal generators in two dimensions as given below:
\begin{align}
    \left[J^{\pm}_{{p}{q}},J^{\pm}_{{r}{s}}\right]&=-i\sqrt{2}\left(g_{{q}{s}}J^{\pm}_{{p}{r}}-g_{{q}{r}}J^{\pm}_{{p}{s}}+g_{{p}{r}}J^{\pm}_{{q}{s}}-g_{{p}{s}}J^{\pm}_{{q}{r}}\right),\nonumber\\
    \left[J^{\pm}_{{p}{q}},J^{\mp}_{{r}{s}}\right]&=0,
\end{align}
which is explicitly given in the Table. \ref{tabelinterpolationlfd}.
We find that $SO(2,1+1)$ splits into a direct sum of two identical algebras:
\begin{align}
\label{LFSO(2,1)}
    SO(2,1+1)\simeq SO(2,1)\oplus SO(2,1)
\end{align}
which is equivalent to Witt algebra in Minkowski space, as we discuss in the next section.

\section{Interpolating  Witt-type Algebra}
\label{sec_Witt-type}
The condition Eq.\eqref{Killing} for invariance under infinitesimal conformal transformations in Euclidean two dimensions gives Cauchy–Riemann equations \cite{Francesco, Blumenhagen}. In general, conformal generators in Euclidean two dimensions are infinity dimensional, which are given by $l_n=-z^{n+1}\partial_{z},~\Bar{l}_n=-\Bar{z}^{n+1}\partial_{\Bar{z}}$, where $z=x^t+ix^3,~ \Bar{z}=x^t-ix^3$, but the globally defined conformal transformations on the Riemann sphere are generated by $l_{-1}$, $l_{0}$, $l_{+1}$, $\Bar{l}_{-1}$, $\Bar{l}_{0}$, and $\Bar{l}_{+1}$. These generators obey the Witt algebra in Euclidean complex space~ \cite{Francesco, Blumenhagen}. The corresponding generators in Minkowski spacetime can be attained by doing a Wick rotation from Euclidean to Minkowski space: $x^t\longrightarrow -ix^{0}$ and $x^3\longrightarrow x^{3}$. We extend our interpolation to a more general Witt-type algebra and introduce the interpolating generators, which read
\begin{align}
    l^{\hat{+}}_n&=(-i\sqrt{2})^{n-2}\left[(c+s)\left(x^{+}\right)^{n+1}\partial_{+}+(c-s)\left(x^{+}\right)^{n+1}\partial_{-}\right],\\
    l^{\hat{-}}_n&=(-i\sqrt{2})^{n-2}\left[(s-c)\left(x^{-}\right)^{n+1}\partial_{+}+(c+s)\left(x^{-}\right)^{n+1}\partial_{-}\right],
\end{align} 
where $s=\sin{\delta}$ and $c=\cos{\delta}$. Then the full Witt-type algebra in the interpolating form reads
\begin{align}
    [l^{\hat{+}}_m,l^{\hat{+}}_n]&=\frac{(m-n)}{\sqrt{2}}[(c^{3}+3cs^{2})l^{\hat{+}}_{m+n}+(s^{3}-sc^{2})l^{\hat{-}}_{m+n}],\\
    [l^{\hat{-}}_m,l^{\hat{-}}_n]&=\frac{(m-n)}{\sqrt{2}}[(c^{3}-cs^{2})l^{\hat{+}}_{m+n}+(s^{3}+3sc^{2})l^{\hat{-}}_{m+n}],\\
    [l^{\hat{+}}_m,l^{\hat{-}}_n]&=\frac{(m-n)}{\sqrt{2}}[(s^{3}-sc^{2})l^{\hat{+}}_{m+n}+(c^{3}-cs^{2})l^{\hat{-}}_{m+n}].
\end{align}
The globally defined conformal transformations on the Riemann sphere are generated by $l^{\hat{\pm}}_{-1}$, $l^{\hat{\pm}}_{0}$, and $l^{\hat{\pm}}_{+1}$, then the explicit commutations read
\begin{subequations}
\begin{align}
    [l^{\hat{+}}_{-1},l^{\hat{+}}_{0}]&=-\frac{1}{\sqrt{2}}[(c^{3}+3cs^{2})l^{\hat{+}}_{-1}+(s^{3}-sc^{2})l^{\hat{-}}_{-1}]\\
    [l^{\hat{+}}_{0},l^{\hat{+}}_{1}]&=-\frac{1}{\sqrt{2}}[(c^{3}+3cs^{2})l^{\hat{+}}_{1}+(s^{3}-sc^{2})l^{\hat{-}}_{1}]\\
    [l^{\hat{+}}_{1},l^{\hat{+}}_{-1}]&=\sqrt{2}[(c^{3}+3cs^{2})l^{\hat{+}}_{0}+(s^{3}-sc^{2})l^{\hat{-}}_{0}]\\
    [l^{\hat{-}}_{-1},l^{\hat{-}}_{0}]&=-\frac{1}{\sqrt{2}}[(c^{3}-cs^{2})l^{\hat{+}}_{-1}+(s^{3}+3sc^{2})l^{\hat{-}}_{-1}]\\
    [l^{\hat{-}}_{0},l^{\hat{-}}_{1}]&=-\frac{1}{\sqrt{2}}[(c^{3}-cs^{2})l^{\hat{+}}_{1}+(s^{3}+3sc^{2})l^{\hat{-}}_{1}]\\
    [l^{\hat{-}}_{1},l^{\hat{-}}_{-1}]&=\sqrt{2}[(c^{3}-cs^{2})l^{\hat{+}}_{0}+(s^{3}+3sc^{2})l^{\hat{-}}_{0}]\\
    [l^{\hat{+}}_{-1},l^{\hat{-}}_{0}]&=-\frac{1}{\sqrt{2}}[(s^{3}-sc^{2})l^{\hat{+}}_{-1}+(c^{3}-cs^{2})l^{\hat{-}}_{-1}]\\
    [l^{\hat{+}}_{0},l^{\hat{-}}_{1}]&=-\frac{1}{\sqrt{2}}[(s^{3}-sc^{2})l^{\hat{+}}_{1}+(c^{3}-cs^{2})l^{\hat{-}}_{1}]\\
    [l^{\hat{+}}_{1},l^{\hat{-}}_{-1}]&=\sqrt{2}[(s^{3}-sc^{2})l^{\hat{+}}_{0}+(c^{3}-cs^{2})l^{\hat{-}}_{0}]\\
    [l^{\hat{-}}_{-1},l^{\hat{+}}_{0}]&=-\frac{1}{\sqrt{2}}[(s^{3}-sc^{2})l^{\hat{+}}_{-1}+(c^{3}-cs^{2})l^{\hat{-}}_{-1}]\\
    [l^{\hat{-}}_{0},l^{\hat{+}}_{1}]&=-\frac{1}{\sqrt{2}}[(s^{3}-sc^{2})l^{\hat{+}}_{1}+(c^{3}-cs^{2})l^{\hat{-}}_{1}]\\
    [l^{\hat{-}}_{1},l^{\hat{+}}_{-1}]&=\sqrt{2}[(s^{3}-sc^{2})l^{\hat{+}}_{0}+(c^{3}-cs^{2})l^{\hat{-}}_{0}]
\end{align}
\label{interpolating-Witt}
\end{subequations}
where
\begin{align}
    l^{\hat{+}}_{-1}&=-\frac{(P_{0}\cos{\delta}+P_{3}\sin{\delta})}{\sqrt{2}} &&l^{\hat{-}}_{-1}=-\frac{(P_{0}\sin{\delta}-P_{3}\cos{\delta})}{\sqrt{2}}\\
    l^{\hat{+}}_{0}&=i\frac{(D\cos{\delta}+K_{3}\sin{\delta})}{\sqrt{2}} &&l^{\hat{-}}_{0}=i\frac{(D\sin{\delta}-K_{3}\cos{\delta})}{\sqrt{2}}\\
    l^{\hat{+}}_{1}&=\frac{(\mathfrak{K}_{0}\cos{\delta}-\mathfrak{K}_{3}\sin{\delta})}{\sqrt{2}} &&l^{\hat{-}}_{1}=\frac{(\mathfrak{K}_{0}\sin{\delta}+\mathfrak{K}_{3}\cos{\delta})}{\sqrt{2}}.
\end{align}
These results of interpolating Witt-type algebra summarized in Eqs.(\ref{interpolating-Witt}a-l) are equivalent to Table~\ref{tabel1+1interpolationlfd} taking into account the notations of 
$\mathbb{C} = c^2 -s^2$ and $\mathbb{S} = 2 cs$. 
In the Instant form limit $s\rightarrow0,~c\rightarrow1$, we have  $l^{\hat{+}}_{-1}\rightarrow l^{0}_{-1}=-\frac{P_{0}}{\sqrt{2}}$, $l^{\hat{-}}_{-1}\rightarrow l^{3}_{-1}= \frac{P_{3}}{\sqrt{2}}$, $l^{\hat{+}}_{0}\rightarrow l^{0}_{0}=i\frac{D}{\sqrt{2}}$, 
$l^{\hat{-}}_{0}\rightarrow l^{3}_{0}= -i\frac{K_{3}}{\sqrt{2}}$, $l^{\hat{+}}_{1}\rightarrow l^{0}_{1}= \frac{\mathfrak{K}_{0}}{\sqrt{2}}$, and $l^{\hat{-}}_{1}\rightarrow l^{3}_{1}= \frac{\mathfrak{K}_{3}}{\sqrt{2}}$, the commutation relations read
 \begin{align}
    [l^{0}_m,l^{0}_n]&=\frac{(m-n)}{\sqrt{2}}l^{0}_{m+n},\\
    [l^{3}_m,l^{3}_n]&=\frac{(m-n)}{\sqrt{2}}l^{0}_{m+n},\\
    [l^{0}_m,l^{3}_n]&=\frac{(m-n)}{\sqrt{2}}l^{3}_{m+n},\\
    [l^{3}_m,l^{0}_n]&=\frac{(m-n)}{\sqrt{2}}l^{3}_{m+n},
\end{align}
which reproduces the full commutation table mentioned in Table~\ref {tabelinterpolationifd}.
Likewise, 
in the light-front limit $s\rightarrow\frac{1}{\sqrt{2}},~c\rightarrow\frac{1}{\sqrt{2}}$, we have $ l^{\hat{\pm}}_{-1}\rightarrow l^{\pm}_{-1}=-\frac{P_{\pm}}{\sqrt{2}}$, $l^{\hat{\pm}}_{0}\rightarrow l^{\pm}_{0}=i\frac{D_{\pm}}{\sqrt{2}}$, and $l^{\hat{\pm}}_{1}\rightarrow l^{\pm}_{1}= \frac{\mathfrak{K}_{\pm}}{\sqrt{2}}$, then the commutation relations reproduce the known Witt-type algebra,
\begin{align}
    [l^{\pm}_m,l^{\pm}_n]&=(m-n)l^{\pm}_{m+n},\\
    [l^{\pm}_m,l^{\mp}_n]&=0,
\end{align}
which reproduces the full commutation table mentioned in Table~\ref {tabelinterpolationlfd}. This algebra implies the $SO(2,1+1)$ splits into a direct sum of two identical algebras $SO(2,1)\oplus SO(2,1)$ as given by Eq.(\ref{LFSO(2,1)}).

\section{\texorpdfstring{$4\times4$}{TEXT} Projective Spacetime Representations}
\label{sec:projectivespace}
In order to characterize the individual properties of the six conformal generators, namely, which generators are kinematic or dynamic, we may correspond the projective spacetime representations 
given by Eq.(\ref{Jab}) with
the interpolating $(1+1)$ dimensional conformal algebra
discussed in the previous sections, Secs.\ref{sec:conformal} and \ref{sec_Witt-type}. 
Kinematic generators leave the time defined in the given form of the dynamics invariant.
Consequently, the individual time-ordered amplitudes in the respective time-ordered processes are invariant under the transformations provided by kinematic generators \cite{Ji2001}.
Figuring out which generator is kinematic in what form of dynamics is useful as the more kinematic generators save the correspondingly more dynamic efforts in solving the relativistic quantum field theories, such as QED and QCD~ \cite{ji2023relativistic}. As discussed in Sec.\ref{sec:conformal}, 
LFD saves dynamic efforts in solving the relativistic quantum field theoretic problems due to the character change of the longitudinal boost operator $K_3$ from dynamic in IFD to kinematic in LFD, as exemplified in 'tHooft model computations  \cite{THOOFT1974461, Ji2021QCD, ji2023relativistic}.

 To find the $4\times4$ projective spacetime representations in the interpolating form, it is convenient to use the 
 $4\times4$ transformation matrix given by
\begin{align}
    (\mathcal{R}_{\hat{a}}^{b})_{4\times4}=(\mathcal{R}_{\hat{a}}^{b})^T_{4\times4}=\begin{pmatrix}
    1&0&0&0\\
    0&1&0&0\\
    0&0&\cos{\delta}&\sin{\delta}\\
    0&0&\sin{\delta}&-\cos{\delta}
    \end{pmatrix}_{4\times4}
\end{align}
and apply it to Eq.\eqref{Jab}. 
Then, in the interpolating form, $J_{\hat{a}\hat{b}}$ becomes $J_{\hat{a}\hat{b}}=\mathcal{R}_{\hat{a}}^{~{c}}J_{cd}\mathcal{R}_{~\hat{b}}^{{d}}$, that is
\begin{align}
    J_{\hat{a}\hat{b}}&=\begin{pmatrix}
    0&-D&-\frac{\mathfrak{K}^{\hat{+}}}{\sqrt{2}}&-\frac{\mathfrak{K}^{\hat{-}}}{\sqrt{2}}\\
    D&0&\frac{P_{\hat{+}}}{\sqrt{2}}&\frac{P_{\hat{-}}}{\sqrt{2}}\\
    \frac{\mathfrak{K}^{\hat{+}}}{\sqrt{2}}&-\frac{P_{\hat{+}}}{\sqrt{2}}&0  & -{K}_{3}\\
    \frac{\mathfrak{K}^{\hat{-}}}{\sqrt{2}}&-\frac{P_{\hat{-}}}{\sqrt{2}}&{K}_{3}  & 0
  \end{pmatrix}_{4\times4}\label{Jhat+hat-},
\end{align}
where, $P_{\hat{+}}=P_0\cos{\delta}+P_3\sin{\delta}$, $P_{\hat{-}}=P_0\sin{\delta}-P_3\cos{\delta}$, $
\mathfrak{K}^{\hat{+}}=\mathfrak{K}_0\cos{\delta}+\mathfrak{K}_3\sin{\delta}$, $\mathfrak{K}^{\hat{-}}=\mathfrak{K}_0\sin{\delta}-\mathfrak{K}_3\cos{\delta}$, $D_{\hat{+}}=D\cos\delta+K_{3}\sin\delta$, 
and $D_{\hat{-}}=D\sin\delta-K_{3}\cos\delta$. The extension to the $6 \times 6$ projective spacetime representations corresponding to the interpolating (3+1) dimensional conformal algebra can be made straightforwardly, as the perpendicular components are the same with or without the interpolation.

The interpolation in projective spacetime naturally produces superscript special conformal generators  ($\mathfrak{K}^{\hat{\pm}}$), which are related to subscript special conformal generators as $\mathfrak{K}^{\hat{\pm}}=g^{\hat{\pm}\hat{\pm}}\mathfrak{K}_{\hat{\pm}}+g^{\hat{\pm}\hat{\mp}}\mathfrak{K}_{\hat{\mp}}$, where the metric is defined in Eq.\eqref{eqn:g_munu_interpolation}, explicitly,
\begin{align}
    \begin{pmatrix}
        \mathfrak{K}^{\hat{+}}\\
        \mathfrak{K}^{\hat{-}}
    \end{pmatrix}=&\begin{pmatrix}
        \mathbb{C}&\mathbb{S}\\
        \mathbb{S}&-\mathbb{C}
    \end{pmatrix}\begin{pmatrix}
        \mathfrak{K}_{\hat{+}}\\
        \mathfrak{K}_{\hat{-}}
    \end{pmatrix}.
\end{align}
In the IFD limit $\delta\rightarrow0$ (or $\mathbb{S}\rightarrow0$ \& $\mathbb{C}\rightarrow1$), we have $P_{\hat{+}}\rightarrow P_{0}$, $P_{\hat{-}}\rightarrow -P_{3}$, $\mathfrak{K}^{\hat{+}}\rightarrow \mathfrak{K}_{0}$, $\mathfrak{K}^{\hat{-}}\rightarrow -\mathfrak{K}_{3}$, $D_{\hat{+}}\rightarrow D$, and $D_{\hat{-}}\rightarrow -K_{3}$. Likewise, in the LFD limit $\delta\rightarrow\frac{\pi}{4}$ (or $\mathbb{S}\rightarrow1$ \& $\mathbb{C}\rightarrow0$), we have $P_{\hat{\pm}}\rightarrow P_{\pm}=\frac{P_0\pm P_3}{\sqrt{2}}$, $\mathfrak{K}^{\hat{\pm}}\rightarrow \mathfrak{K}_{\mp}=\frac{\mathfrak{K}_0\pm \mathfrak{K}_3}{\sqrt{2}}$, and $D_{\hat{\pm}}\rightarrow D_{\pm}=\frac{D \pm K_3}{\sqrt{2}}$. Then the simplified conformal algebra in projective spacetime interpolation is:
  \begin{align}
      \left[J_{{\hat{a}}{\hat{b}}}J_{{\hat{c}}{\hat{d}}}\right]=-i\left(g_{{\hat{b}}{\hat{d}}}J_{{\hat{a}}{\hat{c}}}-g_{{\hat{b}}{\hat{c}}}J_{{\hat{a}}{\hat{d}}}+g_{{\hat{a}}{\hat{c}}}J_{{\hat{b}}{\hat{d}}}-g_{{\hat{a}}{\hat{d}}}J_{{\hat{b}}{\hat{c}}}\right)\label{simplesrint}
  \end{align}
where, 
\begin{align}
    g_{\hat{a}\hat{b}}&=\begin{pmatrix}
  0&-1&0&0\\
  -1&0&0&0\\
  0&0&\mathbb{C}&\mathbb{S}\\
  0&0&\mathbb{S}&-\mathbb{C}\\
  \end{pmatrix}_{4\times4}\label{metricghat}.
\end{align}
The algebra Eq.\eqref{simplesrint} with the above interpolating $4\times4$ metric will reproduce $_6C_2 = 15$ explicit commutation relations in interpolation between IFD, and LFD mentioned in Table~\ref {tabel1+1interpolationlfd}. The conformal algebra given by Eq.\eqref{simplesrint} in the interpolating form of projective spacetime implies that $J_{\hat{a}\hat{b}}$ can be written as
\begin{align}
    J_{\hat{a}\hat{b}}=i(X_{\hat{a}}\partial_{\hat{b}}-X_{\hat{b}}\partial_{\hat{a}}),
\end{align}
where $\hat{a},\hat{b}\in\{-2,-1,\hat{+},\hat{-}\}$, $\partial_{\hat{a}}\equiv\frac{\partial}{\partial X^{\hat{a}}}$ and $X_{\hat{a}}$ denotes the four-dimensional projective spacetime. With the condition that the lightcone is preserved in the four-dimensional projective spacetime under the transformations generated by $ J_{\hat{a}\hat{b}}$, we write the infinitesimal conformal transformations in four-dimensional projective spacetime as
\begin{align}
R^{\hat{a}}_{~\hat{b}}=g^{\hat{a}}_{~\hat{b}}-\frac{i}{2}\omega^{\hat{c}\hat{d}}(J_{\hat{c}\hat{d}})^{\hat{a}}_{~\hat{b}},\label{R_conformalTransformations}
\end{align}
where $\omega^{\hat{a}\hat{b}}=-\omega^{\hat{b}\hat{a}}$. Here, we may regard $R^{\hat{a}}_{~\hat{b}}$ as the expansion of $e^{-\frac{i}{2}\omega^{\hat{c}\hat{d}}(J_{\hat{c}\hat{d}})^{\hat{a}}_{~\hat{b}}}$ up to the first order of $\omega$ as one can see in the expansion given by 
$e^{-\frac{i}{2}\omega^{\hat{c}\hat{d}}(J_{\hat{c}\hat{d}})^{\hat{a}}_{~\hat{b}}} \approx  g^{\hat{a}}_{~\hat{b}}-\frac{i}{2}\omega^{\hat{c}\hat{d}}(J_{\hat{c}\hat{d}})^{\hat{a}}_{~\hat{b}}+\mathcal{O}(\omega^2)$.
The generator representation $(J_{\hat{c}\hat{d}})^{\hat{a}}_{~\hat{b}}$ can be obtained by
\begin{align}
    (J_{\hat{c}\hat{d}})^{\hat{a}}_{~\hat{b}}=(J_{\hat{c}\hat{d}})^{\hat{a}\hat{f}}g_{\hat{f}\hat{b}},
\end{align}
where $(J_{{\hat{c}}{\hat{d}}})^{{\hat{a}}{\hat{b}}}=i(g_{\hat{c}}^{\hat{a}} g_{\hat{d}}^{\hat{b}}-g_{\hat{c}}^{\hat{b}} g_{\hat{d}}^{\hat{a}})$. The representation matrices of conformal generators are defined by taking the first  index to be a superscript and the second subscript:

\begin{eqnarray}
    &&(D_{\hat{+}})^{\hat{a}}_{~\hat{b}}\equiv -\cos{\delta}(J_{-2-1})^{\hat{a}}_{~\hat{b}}-\sin{\delta}(J_{\hat{+}\hat{-}})^{\hat{a}}_{~\hat{b}}~;\nonumber\\
    &&(D_{\hat{-}})^{\hat{a}}_{~\hat{b}}\equiv -\sin{\delta}(J_{-2-1})^{\hat{a}}_{~\hat{b}}+\cos{\delta}(J_{\hat{+}\hat{-}})^{\hat{a}}_{~\hat{b}}~;\nonumber\\
    &&\frac{-(\mathfrak{K}^{\hat{\mu}})^{\hat{a}}_{~\hat{b}}}{\sqrt{2}}\equiv (J_{-2\hat{\mu}})^{\hat{a}}_{~\hat{b}}~;
    \frac{(P_{\hat{\mu}})^{\hat{a}}_{~\hat{b}}}{\sqrt{2}}\equiv (J_{-1\hat{\mu}})^{\hat{a}}_{~\hat{b}}~.
    \label{implicit6x6}
\end{eqnarray}

The explicit $4\times4$ matrix representations are then given by 
\begin{align}
    D_{\hat{+}}&=\begin{pmatrix}
        -i\cos{(\delta)}&0&0&0\\
        0&i\cos{(\delta)}&0&0\\
        0&0&i\mathbb{S}\sin{(\delta)}&-i\mathbb{C}\sin{(\delta)}\\
        0&0&-i\mathbb{C}\sin{(\delta)}&-i\mathbb{S}\sin{(\delta)}
    \end{pmatrix};\nonumber\\
    D_{\hat{-}}&=\begin{pmatrix}
        -i\sin{(\delta)}&0&0&0\\
        0&i\sin{(\delta)}&0&0\\
        0&0&-i\mathbb{S}\cos{(\delta)}&i\mathbb{C}\cos{(\delta)}\\
        0&0&i\mathbb{C}\cos{(\delta)}&i\mathbb{S}\cos{(\delta)}
    \end{pmatrix};\nonumber
\end{align}
\begin{align}
    \mathfrak{K}^{\hat{+}}&=\sqrt{2}\begin{pmatrix}
        0&0&0&0\\
        0&0&i&0\\
        i\mathbb{C}&0&0&0\\
        i\mathbb{S}&0&0&0
    \end{pmatrix};~\mathfrak{K}^{\hat{-}}=\sqrt{2}\begin{pmatrix}
        0&0&0&0\\
        0&0&0&i\\
        i\mathbb{S}&0&0&0\\
        -i\mathbb{C}&0&0&0
    \end{pmatrix};\nonumber
    \end{align}
    \begin{align}
     P_{\hat{+}}&=\sqrt{2}\begin{pmatrix}
        0&0&-i&0\\
        0&0&0&0\\
        0&-i\mathbb{C}&0&0\\
        0&-i\mathbb{S}&0&0
    \end{pmatrix};~P_{\hat{-}}=\sqrt{2}\begin{pmatrix}
        0&0&0&-i\\
        0&0&0&0\\
        0&-i\mathbb{S}&0&0\\
        0&i\mathbb{C}&0&0
    \end{pmatrix}.
    \label{matrix-rep-conformal-generators}
    \end{align}
These explicit projective spacetime matrices satisfy the full conformal algebra Eq.\eqref{simplesrint} in interpolation. 

Operating the exponentiated conformal transformations $e^{-\frac{i}{2}\omega^{\hat{c}\hat{d}}(J_{\hat{c}\hat{d}})^{\hat{a}}_{~\hat{b}}}$ on $X_{\hat{b}}$, we get the transformed four-dimensional projective spacetime $X^{\prime}_{\hat{a}}$ as 
\begin{align}
\label{exponential-conformal-tranf}
  X^{\prime}_{\hat{a}}&=e^{-\frac{i}{2}\omega^{\hat{c}\hat{d}}(J_{\hat{c}\hat{d}})^{\hat{b}}_{~\hat{a}}}X_{\hat{b}},
\end{align}
and the inner product
\begin{align}
{X^{\prime}}_{\hat{a}}{X^{\prime}}^{\hat{a}} &= 
e^{-\frac{i}{2}\omega^{\hat{c}\hat{d}}(J_{\hat{c}\hat{d}})^{\hat{b}}_{~\hat{a}}}X_{\hat{b}}e^{-\frac{i}{2}\omega^{\hat{c}\hat{d}}(J_{\hat{c}\hat{d}})^{\hat{a}}_{~\hat{b}}}X^{\hat{b}}\nonumber\\
&= e^{-\frac{i}{2}\omega^{\hat{c}\hat{d}}(J_{\hat{c}\hat{d}})^{\hat{b}}_{~\hat{a}}}e^{-\frac{i}{2}\omega^{\hat{c}\hat{d}}(J_{\hat{c}\hat{d}})^{\hat{a}}_{~\hat{b}}}X_{\hat{b}}X^{\hat{b}}\nonumber\\
&= e^{-\frac{i}{2}\omega^{\hat{c}\hat{d}}((J_{\hat{c}\hat{d}})^{\hat{b}}_{~\hat{a}}+(J_{\hat{c}\hat{d}})^{\hat{a}}_{~\hat{b}})}X_{\hat{b}}X^{\hat{b}}\nonumber\\
&= X_{\hat{b}}X^{\hat{b}},
\end{align}
where we used $(J_{{\hat{c}}{\hat{d}}})^{\hat{a}}_{~\hat{b}}=i(g_{\hat{c}}^{\hat{a}} g_{{\hat{d}}{\hat{b}}}-g_{{\hat{c}}{\hat{b}}} g_{\hat{d}}^{\hat{a}})$ to obtain $(J_{\hat{c}\hat{d}})^{\hat{a}}_{~\hat{b}}=-(J_{\hat{c}\hat{d}})^{\hat{b}}_{~\hat{a}}$.
This shows that the conformal transformations, even in the interpolating form, preserve the angles and the ratios of lengths of the two vectors multiplied by each other before and after the transformations. Now, we find that the assignment of the following projective four-dimensional vector given by 
\begin{align}
    {X}_{-1}&=\frac{-\lambda}{\sqrt{2}}, \nonumber\\
    {X}_{-2}&=\frac{-\lambda}{\sqrt{2}}(x^{\hat{\mu}}x_{\hat{\mu}}), \nonumber \\
     {X}_{\hat{\mu}}&=\lambda x_{\hat{\mu}},\label{X-mu}
\end{align}
provides the lightcone of the projective spacetime, i.e., ${X}_{\hat{a}}{X}^{\hat{a}}=0$, as explicitly shown below: 
\begin{align}
    {X}_{\hat{a}}{X}^{\hat{a}}&={X}_{-2}{X}^{-2}+{X}_{-1}{X}^{-1}+{X}_{\hat{\mu}}{X}^{\hat{\mu}} \nonumber \\
    &=-2{X}_{-2}{X}_{-1}+{X}^{\hat{\mu}}{X}_{\hat{\mu}}\nonumber \\
     &=-2\frac{-\lambda}{\sqrt{2}}(x^{\hat{\mu}}x_{\hat{\mu}})\frac{-\lambda}{\sqrt{2}}+\lambda^2{x}^{\hat{\mu}}{x}_{\hat{\mu}}\nonumber \\
     &=-\lambda^2{x}^{\hat{\mu}}{x}_{\hat{\mu}}+\lambda^2{x}^{\hat{\mu}}{x}_{\hat{\mu}}\nonumber \\
     &=0,
\end{align}
where ${X}^{-1}=-{X}_{-2}$, ${X}^{-2}=-{X}_{-1}$ and $X^{\hat \mu}=g^{\hat{\mu}\hat{\nu}}X_{\hat{\nu}}$ due to the metric
given by Eq.(\ref{metricghat}).
From Eq.\eqref{X-mu}, we can then find 
the dictionary connecting four-dimensional projective spacetime $X_{\hat{a}}$ and two dimensional spacetime $x_{\hat{\mu}}$ which reads
\begin{align}
    x_{\hat{\mu}}&=-\frac{1}{\sqrt{2}}\frac{{X}_{\hat{\mu}}}{ {X}_{-1}}.
\end{align}
Corresponding to $x_{\hat{\mu}}$, we have  
\begin{align}
    x^{\hat{\mu}}&=-\frac{1}{\sqrt{2}}\frac{{X}^{\hat{\mu}}}{ {X}_{-1}}.
\end{align}
Then, the inverse spacetime of $x^{\hat{\mu}}$ is given by $\frac{x_{\hat{\mu}}}{x^2}=-\frac{1}{\sqrt{2}}\frac{{X}_{\hat{\mu}}}{{X}_{-2}}$, with $x^2=\frac{{X}_{-2}}{{X}_{-1}}$. Under the transformation given by Eq.(\ref{exponential-conformal-tranf}), we compute the transformation of interpolating time given by
\begin{align}
    x^{\hat{+}\prime}&=-\frac{1}{\sqrt{2}}\frac{{X}^{\hat{+}\prime}}{ {X}^{\prime}_{-1}},
\end{align}
for each and every interpolating conformal operation according to Eq.(\ref{exponential-conformal-tranf}). We summarized the results of our computation in Table~\ref{tablexprime}.
From these results, we can determine which generators leave the interpolating time intact, i.e, $x^{\hat{+}\prime} = x^{\hat{+}}$ modulo a scale factor, before and after the transformation. These generators, which result in $x^{\hat{+}\prime}=x^{\hat{+}}$ modulo a scale factor, are the kinematic generators as we discussed earlier. For $0\leq \delta < \pi/4$, we find $P_{\hat -} = \sqrt{2}J_{-1\hat{-}}$ and $D=\left(D_{\hat{+}}c+D_{\hat{-}}s\right)=J_{-1 -2}$ are the only two kinematic generators which yield $x^{\hat{+}\prime}=x^{\hat{+}}$ modulo a scale factor.
The scale factor for the interpolating longitudinal momentum $P_{\hat{-}}$ is one, while the scale factor for the dilatation $D$ is $e^{-\alpha}$ with $\alpha=\omega^{-1-2}$ in Eq.(\ref{exponential-conformal-tranf}) as one can find straightforwardly from the $4\times4$ matrix representations of $D_{\hat{+}}$ and $D_{\hat{-}}$ in Eq.(\ref{matrix-rep-conformal-generators}). 
Taking the respective limit of the interpolating parameter $\delta$, namely,  
$\delta \to 0$ and $\delta \to \pi/4$, we find the corresponding number of kinematic generators in the IFD and the LFD, respectively. 
\begin{widetext}
\begin{center}
\begin{table}[h!]
        \scalebox{1}{ 
        \begin{tabular}{|c||c|c|}
        \hline
             \rule{0pt}{16pt} Generators & $x^{\hat{+}\prime}$\\
             \hline
             \hline
            \rule{0pt}{16pt} $\mathfrak{K}^{\hat{+}}$ & $x^{\hat{+}\prime}=\frac{x^{\hat{+}}-b^{\hat{+}}(x)^2}{1-2\mathbb{C}b^{\hat{+}}x^{\hat{+}}-2\mathbb{S}b^{\hat{+}}x^{\hat{-}}+\mathbb{C}(b^{\hat{+}})^2(x)^2}$\\
             \hline
            \rule{0pt}{16pt} $\mathfrak{K}^{\hat{-}}$ & $x^{\hat{+}\prime}=\frac{x^{\hat{+}}}{1+2\mathbb{C}b^{\hat{-}}x^{\hat{-}}-2\mathbb{S}b^{\hat{-}}x^{\hat{+}}-\mathbb{C}(b^{\hat{-}})^2(x)^2}$ \\
             \hline
             \rule{0pt}{16pt} $P_{\hat{+}}$& $x^{\hat{+}\prime}=x^{\hat{+}}+c^{\hat{+}} $\\
             \hline
             \rule{0pt}{16pt} $P_{\hat{-}}$& $x^{\hat{+}\prime}=x^{\hat{+}}$  \\
             \hline
             \rule{0pt}{16pt} $D_{\hat{+}}$ & $x^{\hat{+}\prime} = e^{-\alpha\cos{(\delta)}}\left[\left(\cosh[\alpha\sin{(\delta)}]-\mathbb{S}\sinh[\alpha\sin{(\delta)}]\right)x^{\hat{+}}+\mathbb{C}\sinh[\alpha\sin{(\delta)}] x^{\hat{-}}\right]$ \\
             \hline
             \rule{0pt}{16pt} $D_{\hat{-}}$ & $x^{\hat{+}\prime} = e^{-\alpha\sin{(\delta)}}\left[\left(\cosh[\alpha\cos{(\delta)}]+\mathbb{S} \sinh[\alpha\cos{(\delta)}]\right)x^{\hat{+}} -\mathbb{C} \sinh[\alpha\cos{(\delta)}]x^{\hat{-}}\right]$ \\
             \hline
        \end{tabular}}
        \caption{Transformation of interpolating time under each conformal generator in $(1+1)$ dimensions. The parameters $\omega^{\hat{c}\hat{d}}$ in Eq.(\ref{exponential-conformal-tranf}) surviving in this transformation are denoted by  
$b^{\hat{\pm}}=\omega^{\hat{\pm},-2},c^{\hat{+}}=\omega^{-1\hat{+}}$ and $\alpha=\omega^{-1-2}=\omega^{\hat{-}\hat{+}}$, respectively, while the interpolating parameter between IFD and LFD is denoted by $\delta$.}
        \label{tablexprime}
    \end{table}
    \end{center}

\begin{center}
\begin{table}[!htb]
        \begin{minipage}{.5\linewidth}
        \begin{tabular}{|c||c|c|}
        \hline
             \rule{0pt}{16pt} Generators & $x^{{0}\prime}$\\
             \hline
             \hline
            \rule{0pt}{16pt} $\mathfrak{K}_{{0}}$ & $x^{{0}\prime}=\frac{x^{{0}}-b^{{0}}(x)^2}{1-2b^{{0}}x^{{0}}+(b^{{0}})^2(x)^2}$\\
             \hline
            \rule{0pt}{16pt} $-\mathfrak{K}_{{3}}$ & $x^{{0}\prime}=\frac{x^{{0}}}{1+2b^{{3}}x^{{3}}-(b^{{3}})^2(x)^2}$ \\
             \hline
             \rule{0pt}{16pt} $P_{{0}}$& $x^{{0}\prime}=x^{{0}}+c^{{0}} $\\
             \hline
             \rule{0pt}{16pt} $-P_{{3}}$& $x^{{0}\prime}=x^{{0}}$  \\
             \hline
             \rule{0pt}{16pt} $D$ & $x^{{0}\prime} = e^{-\alpha}x^{0}$ \\
             \hline
             \rule{0pt}{16pt} $-K_{3}$ & $x^{{0}\prime} = \cosh[\alpha]x^{0}+\sinh[\alpha] x^{3}$ \\
             \hline
        \end{tabular}
        \caption{Transformation of the IFD time in $(1+1)$d}
        \label{tablex0prime}
        \end{minipage}%
    \begin{minipage}{.5\linewidth}
        \begin{tabular}{|c||c|c|}
        \hline
             \rule{0pt}{16pt} Generators & $x^{{+}\prime}$\\
             \hline
             \hline
            \rule{0pt}{16pt} $\mathfrak{K}_{{-}}$ & $x^{{+}\prime}=x^{+}$\\
             \hline
            \rule{0pt}{16pt} $\mathfrak{K}_{{+}}$ & $x^{{+}\prime}=\frac{x^{{+}}}{1-2b^{{-}}x^{{+}}}$ \\
             \hline
             \rule{0pt}{16pt} $P_{{+}}$& $x^{{+}\prime}=x^{{+}}+c^{{+}} $\\
             \hline
             \rule{0pt}{16pt} $P_{{-}}$& $x^{{+}\prime}=x^{{+}}$  \\
             \hline
             \rule{0pt}{16pt} $D_{{+}}$ & $x^{{+}\prime} = e^{-\sqrt{2}\alpha}x^{{+}}$ \\
             \hline
             \rule{0pt}{16pt} $D_{{-}}$ & $x^{{+}\prime} = x^{{+}}$ \\
             \hline
        \end{tabular}
        \caption{Transformation of LFD time in $(1+1)$d}
        \label{tablexplusprime}
          \end{minipage} 
    \end{table}
    \end{center}

\begin{table}[h!]
    \begin{ruledtabular}
       \begin{tabular}{lcc}
	Interpolation angle & Kinematic & Dynamic \\
	\hline
	\rule{0pt}{3ex} $\delta=0$ & $P_{3}$, $D$ & $ K_{3}$, $P_{0}$, $\mathfrak{K}_{{0}}$, $\mathfrak{K}_{{3}}$\\
	$0<\delta<\pi/4$ &  $P_{\mT}$, $D=\left(D_{\hat{+}}c+D_{\hat{-}}s\right)$ &  $K_3=\left(D_{\hat{+}}s-D_{\hat{-}}c\right)$,  $P_{\pT}$, $\mathfrak{K}^{\hat{+}}$, $\mathfrak{K}^{\hat{-}}$\\
	$\delta=\pi/4$ & $P_{-}$, $D_{+}=\frac{(D+K_3)}{\sqrt{2}} $, $D_{-}=\frac{(D-K_3)}{\sqrt{2}}$, $\mathfrak{K}_{{-}}$& $P_{+}$, $\mathfrak{K}_{{+}}$\\
      \end{tabular}
    \end{ruledtabular}
    \caption{\label{tab:Kinematic_and_dynamic_generators_for_different_interoplation_angles_conformal1+1}Kinematic and dynamic conformal generators for different interpolation angles in $(1+1)$}
\end{table}
\end{widetext}

In the limit to $\delta=0$, i.e., the IFD ($\mathbb{C}=1;~\mathbb{S}=0$), we find the results shown in Table~\ref{tablex0prime} for the transformed $x^{0\prime}$ 
for each and every (1+1)d conformal generators. As expected from the results shown in Table~\ref{tablexprime} for the interpolating time transformations, we find the two kinematic generators $P^3 = - P_3$ and $D$ corresponding 
to $P_{\hat -} = \sqrt{2}J_{-1\hat{-}}$ and $D=\left(D_{\hat{+}}c+D_{\hat{-}}s\right)=J_{-1 -2}$ for $\delta =0$, respectively.

In the limit to $\delta=\pi/4$, i.e., the LFD ($\mathbb{C}=0;~\mathbb{S}=1$), we find the results shown in Table~\ref{tablexplusprime} for the transformed $x^{+\prime}$ for each and every (1+1)d conformal generators. As expected from the results shown in Table~\ref{tablexprime} for the interpolating time transformations, we find that the generators $P_{-} = \frac{P_{0}-P_{3}}{\sqrt{2}}$ and $D$ corresponding to $P_{\hat -} = \sqrt{2}J_{-1\hat{-}}$ and $D=\left(D_{\hat{+}}c+D_{\hat{-}}s\right)=J_{-1 -2}$ for $\delta =\pi/4$, respectively are kinematic. In addition, the light-front boost $K_3$ corresponding to $K_3=\left(D_{\hat{+}}s-D_{\hat{-}}c\right)=J_{\hat{-}\hat{+}}$ for $\delta =\pi/4$,  becomes kinematic. Therefore the linear combination of $D$ and $K_3$ in light-front, namely  $D_{+}=\frac{(D+K_3)}{\sqrt{2}} $, and $D_{-}=\frac{(D-K_3)}{\sqrt{2}}$ becomes kinematic. 

Finally, the transformations further simplifies for $\mathfrak{K}^{+}=\mathfrak{K}_{-}=\frac{\mathfrak{K}_{0}+\mathfrak{K}_{3}}{\sqrt{2}}$ corresponding to $\mathfrak{K}^{\hat{+}}=\sqrt{2}J_{-2\hat{+}}$ in LFD limit. The light-front time, $x^{+}$ transforms as $ x^{+\prime} =x^{+}$ under special conformal generator $\mathfrak{K}_{-}$. Thus, we get two more kinematic generators $K_3$ and $\mathfrak{K}^{+}$ in LFD ($\delta = \pi/4$) than in IFD ($\delta = 0$) as well as in between ($0 \le \delta <\pi/4$). As discussed earlier, the light-front maximizes the total number of kinematic generators. We found that four out of six generators are kinematic in $(1+1)$d LFD. The set of kinematic and dynamic conformal generators in $(1+1)$ depending on the interpolation angle is summarized in Table~\ref {tab:Kinematic_and_dynamic_generators_for_different_interoplation_angles_conformal1+1}.

\section{Summary and Conclusion}
\label{summary-conclusion}

In this work, we presented the interpolating conformal algebra between IFD and LFD in $(1+1)$ dimensions, 
starting from the lowest dimensions, $(0+1)$ and $(1+0)$, respectively. 
In $(0+1)$ dimensions, we corresponded our discussions based on the Möbius transformation to the formulation provided by de Alfaro, Fubini, and Furlan with the Pauli matrices~\cite{Fubini1976} for the three conformal generators.

In $(1+0)$ dimensions, we discussed the conformal algebra in terms of the ladder operators, utilizing the analogue of one-dimensional simple harmonic oscillator formulation with the creation and annihilation operators. In particular, we provided a physical interpretation of the dilatation operator as another coherent state generator, far different from the translation operator generating Glauber's coherent state, namely the eigenstate of the annihilation operator. It motivates further studies on the different kinds of coherent states, including the special conformal transformation that remains to be explored in future work. Based on the lowest dimensional conformal algebra, we presented the $(1+1)$ dimensional conformal algebra interpolating the two different forms of relativistic dynamics, namely, IFD and LFD. Our work confirms the drastic simplification of the conformal algebra in LFD with the equivalence to Witt algebra in Minkowski space splitting $SO(2,1+1)$ into a direct sum of two identical algebras of $SO(2,1)$ as denoted in Eq.(\ref{LFSO(2,1)}).

We also elaborated the $(1+1)$ dimensional conformal algebra in the interpolating projective spacetime representation and showed that the LFD maximizes the number of kinematic generators, adding the special conformal transformation  generator $\mathfrak{K}^{+}$
as well as the boost generator $K_3$ beyond the typical common kinematic generators of translation and dilatation generators for all forms of the relativistic dynamics for $0 \le \delta \le \pi/4$. 
As discussed in this work, identifying four out of six generators in the conformal algebra as the kinematic generators in LFD appears useful in understanding the utility of LFD for hadron phenomenology, thereby saving considerable dynamic efforts in solving QCD. In this vein, our future work will explore the extension of these interpolation methods to the $(3+1)$  dimensional conformal group. 

As we discussed the correspondence between $(0+1)$d quantum mechanics and AdS$_2$~\cite{Fubini1976} and extended it effectively to the correspondence between $(1+1)$d quantum field theory and AdS$_3$, we look forward to exploring the analysis of the holographic dual correspondence between $(3+1)$d QCD and AdS$_5$, implementing further the interpolation between IFD and LFD. With this analysis, we expect to discuss the effectiveness of the light-front QCD as it maximizes the number of the kinematic generators and provides the link between the first principle QCD and the light-front quark model for the hadron phenomenology. The effectiveness of handling the relativistic interactions among the constituents of the hadrons is highly relevant to the discussion on the advantages that the different forms of relativistic dynamics can offer~\cite{Ji2025}.

\acknowledgments
This work was supported by the U.S. Department of Energy (Grant No. DE-FG02-03ER41260).
The National Energy Research Scientific Computing Center (NERSC) supported by the Office of Science of the U.S. Department of Energy under Contract No. DE-AC02-05CH11231 is also acknowledged. 

\appendix

\section{Dilated vacuum by \texorpdfstring{$D_{3}^{(1+0)}$}{TEXT}} 
\label{Dilated-vacuum}
The new vacuum $\ket{\Omega_{D_{3}}}$ created by $a^{\dagger\prime}_{D^{(1+0)}_{3}}$ in Eq.\eqref{adaggerD1+0} can be derived similarly to the field theory derivation mentioned in Ref. \cite{umezawa1982thermo}. The dilated vacuum by $D_{3}^{(1+0)}$ can be annihilated by $a^{\prime}_{D^{(1+0)}_{3}}$ from Eq.\eqref{aD1+0}, as $ a^{\prime}_{D^{(1+0)}_{3}}\ket{\Omega_{D_{3}}}=0$, which gives, 
\begin{align}
    \ket{\Omega_{D_{3}}}=e^{\frac{\alpha}{2}\left(a^{\dagger 2} - a^2+1\right)}\ket{0}
\end{align}
where, $a\ket{0}=0$. To simplify the exponent, we define $d_{+}=\frac12 a^{\dagger2}$, $d_{-}=\frac12 a^{2}$, and $d_{0}=\frac12\bigl(a^{\dagger}a+\tfrac12\bigr)$, which forms a closed group as $[d_{+},d_{-}]=-2d_{0}$ and $[d_{0},d_{\pm}]=\pm d_{\pm}$. Then
\begin{align}
    \ket{\Omega_{D_{3}}}=e^{\frac{\alpha}{2}}S(\alpha)\ket{0},
\end{align}
where, $S(\alpha)=e^{\alpha(d_{+}-d_{-})}$. The ansatz for $S(\alpha)$, reads
\begin{align}
S(\alpha)=e^{f(\alpha)d_{+}}\;e^{g(\alpha)d_{0}}\;e^{h(\alpha)d_{-}},
\end{align}
with initial conditions $f(0)=g(0)=h(0)=0$. On differentiating the ansatz, and matching the coefficients of $d_{+}$, $d_{-}$, and $d_{0}$, we get the following ordinary differential equation system
\begin{align}
f'-g'f+h'e^{-g}f^{2}&=1,\\
g'-2h'e^{-g}f&=0,\\
h'e^{-g}&=-1.
\end{align}

We also used the Baker–Campbell–Hausdorff formula
\begin{align}
    e^{f d_{+}}d_{0}e^{-f d_{+}}&=d_{0}-f d_{+},\\
e^{g d_{0}}d_{-}e^{-g d_{0}}&=e^{-g}d_{-},\\
e^{f d_{+}}d_{-}e^{-f d_{+}}&=K_{-}-2fd_{0}+f^{2}d_{+}.
\end{align}
The solution for the ordinary differential equation reads 
\begin{align}
    f(\alpha)&=\tanh\alpha,\\
g(\alpha)&=-2\ln(\cosh\alpha),\\
h(\alpha)&=-\tanh\alpha.
\end{align}
Which simplifies the $S(\alpha)$ to 
\begin{align}
    S(\alpha)
   =e^{\tfrac12\tanh\alpha\,a^{\dagger2}} e^{-\ln(\cosh\alpha)\bigl(a^{\dagger}a+\tfrac12\bigr)} e^{-\tfrac12\tanh\alpha\,a^{2}}.
\end{align}

Acting it on vacuum and multiplying by the factor \(e^{\alpha/2}\) gives the final result
\begin{align}
    \ket{\Omega_{D_{3}}}&=\frac{e^{\alpha/2}}{\sqrt{\cosh\alpha}}\;
     e^{\left(\frac{\tanh\alpha}{2}\,a^{\dagger2}\right)}\ket{0}~.
\end{align}

We find the normalization of this new vacuum by computing,
\begin{align}
    \bra{0}&e^{\frac{\tanh\alpha}{2}\,a^{2}} e^{\frac{\tanh\alpha}{2}\,a^{\dagger2}}\ket{0}\nonumber\\
    &=\sum_{n=0}^{\infty}\sum_{k=0}^{\infty}\frac{((\tanh\alpha)/2)^n}{n!}\frac{((\tanh\alpha)/2)^k}{k!}\bra{0}a^{2n} (a^{\dagger2})^k\ket{0}\nonumber\\
    &=\sum_{k=0}^{\infty}\frac{((\tanh\alpha)/2)^k}{k!}\frac{((\tanh\alpha)/2)^k}{k!}\sqrt{(2k)!}\bra{0}a^{(2k)} \ket{2k}\nonumber\\
    &=\sum_{k=0}^{\infty}\frac{((\tanh^2{\alpha})/4)^k}{k!k!}(2k)!\nonumber\\
    &=\cosh{\alpha}
\end{align}
with normalization of vacuum $\braket{0|0}=1$, then
\begin{align}
    \braket{\Omega_{D_{3}}|\Omega_{D_{3}}} =&  \frac{e^{\alpha}}{\cosh\alpha}  \bra{0}e^{\frac{\tanh\alpha}{2}\,a^{2}} e^{\frac{\tanh\alpha}{2}\,a^{\dagger2}}\ket{0}=e^{\alpha}.
\end{align}

Similarly, we find the inner product of the dilated vacuum with
the displaced vacuum by computing
\begin{align}
    \bra{0}&e^{-\frac{c}{\sqrt{2}}a} e^{\frac{\tanh\alpha}{2}\,a^{\dagger2}}\ket{0}\nonumber\\
    &=\sum_{n=0}^{\infty}\sum_{k=0}^{\infty}\frac{(-c/\sqrt{2})^n}{n!}\frac{((\tanh\alpha)/2)^k}{k!}\bra{0}a^n (a^{\dagger2})^k\ket{0}\nonumber\\
    &=\sum_{k=0}^{\infty}\frac{(-c/\sqrt{2})^{(2k)}}{(2k)!}\frac{((\tanh\alpha)/2)^k}{k!}\sqrt{(2k)!}\bra{0}a^{(2k)} \ket{2k}\nonumber\\
    &=e^{\frac{c^2\tanh\alpha}{4}}
\end{align}
then
\begin{align}
    \braket{\Omega_{P_{3}}|\Omega_{D_{3}}} =& \frac{e^{\alpha/2}}{\sqrt{\cosh\alpha}}\; e^{-\frac{c^2}{4}(1-\tanh\alpha)}~.
\end{align}

\bibliography{BibTeXList}

\end{document}